\documentclass[11pt]{article}

\usepackage{graphicx}
\usepackage{epsfig}
\usepackage{cite}
\usepackage{amssymb}
\usepackage{amsmath}
\usepackage{dsfont}
\usepackage{multirow}

\usepackage{url}
\usepackage{hyperref}
\usepackage{amsfonts}

\def \as {\alpha_s}

\def\lsim{\mathrel{
   \rlap{\raise 0.511ex \hbox{$<$}}{\lower 0.511ex \hbox{$\sim$}}}}
\def\gsim{\mathrel{
   \rlap{\raise 0.511ex \hbox{$>$}}{\lower 0.511ex \hbox{$\sim$}}}}

\newcommand {\bq} {\begin{equation}}
\newcommand {\eq} {\end{equation}}

\newcommand{\be}{\begin{equation}}
\newcommand{\ee}{\end{equation}}
\newcommand{\bea}{\begin{eqnarray}}
\newcommand{\eea}{\end{eqnarray}}
\newcommand{\bi}{\begin{itemize}}
\newcommand{\ei}{\end{itemize}}
\newcommand{\ben}{\begin{enumerate}}
\newcommand{\een}{\end{enumerate}}
\newcommand{\la}{\left\langle}
\newcommand{\ra}{\right\rangle}

\newcommand{\lp}{\left(}
\newcommand{\rp}{\right)}

\def\frac#1#2{{{#1}\over {#2}}}
\def\gsim{\mathrel{\rlap{\lower4pt\hbox{\hskip1pt$\sim$}}
    \raise1pt\hbox{$>$}}}         %greater than or approx. symbol
\def\lsim{\mathrel{\rlap{\lower4pt\hbox{\hskip1pt$\sim$}}
    \raise1pt\hbox{$<$}}}         %less than or approx. symbol

\newcommand{\data}{\mathrm{data}}

\newcommand{\cut}{\mathrm{cut}}

\begin{document}
\begin{flushright}
IFUM-948-FT\\
\end{flushright}
\begin{center}
{\Large \bf
Deviations from NLO QCD evolution\\ in inclusive HERA data}
\vspace{0.8cm}

Fabrizio~Caola, Stefano~Forte and 
Juan~Rojo

\vspace{1.cm}
{\it 
Dipartimento di Fisica, Universit\`a di Milano and
INFN, Sezione di Milano,\\ Via Celoria 16, I-20133 Milano, Italy}
%\end{center} 

\bigskip
\bigskip

%\begin{center}
{\bf \large Abstract:}
\end{center}

We search for deviations from next-to-leading order 
QCD evolution in HERA structure
function data. We compare to data
 predictions for structure functions in the small $x$ region, 
obtained by evolving backwards to  low
$Q^2$ the results of a parton fit performed in
the large $Q^2$ region, where fixed-order
perturbative QCD is certainly reliable.
We find evidence for deviations which
 are qualitatively consistent with the behaviour
predicted by small $x$ perturbative resummation, and possibly also by
nonlinear evolution effects, but incompatible with
next-to-next-to-leading order
corrections. 

\clearpage

There are several reasons to
expect that fixed--order next-to-leading perturbative QCD evolution
(NLO DGLAP~\cite{Altarelli:1977zs}  evolution, henceforth)
 might fail to provide an adequate description of experimental data 
for small enough values of Bjorken $x$ and of $Q^2$. These
include the presence of higher order corrections, which are
large at small $x$~\cite{Moch:2004pa,Martin:2009iq}, the possible
impact of all--order resummation of large
small $x$ logs~\cite{resum} or
non linear phenomena which are expected to restore unitarity in the
high energy limit~\cite{GolecBiernat:2008nq}.
On the other hand, available  QCD analysis based on NLO 
DGLAP~\cite{Nadolsky:2008zw,Ball:2009mk,Martin:2009iq} are known to
provide an excellent description of HERA data, which implies that such
deviations, if any, must be small, and perhaps in current
determinations of parton distributions
(PDFs) are partly  
absorbed in the form of a distortion of the PDFs themselves.

The aim of this work is to provide a general 
strategy to quantify potential deviations from NLO DGLAP, and to apply
it to existing HERA data.  The search for
deviations from NLO DGLAP in HERA data has been recently the subject
of intense theoretical and experimental
activity~\cite{Ellis:2008yp,Iancu:2003ge,Albacete:2009fh,Weigert:2005us}.
However, existing studies typically investigate the agreement of the data
with predictions of specific models, rather than trying to provide a 
comparative assessment of the predictions of the models in comparison
to 
NLO DGLAP. Also, the issue of
possible dependence of results on the choice of PDFs
is usually not addressed. Our approach is meant to provide a model
independent assessment of effects beyond NLO DGLAP.

% ---------------- 
\begin{figure}[b]
\centering
\epsfig{file=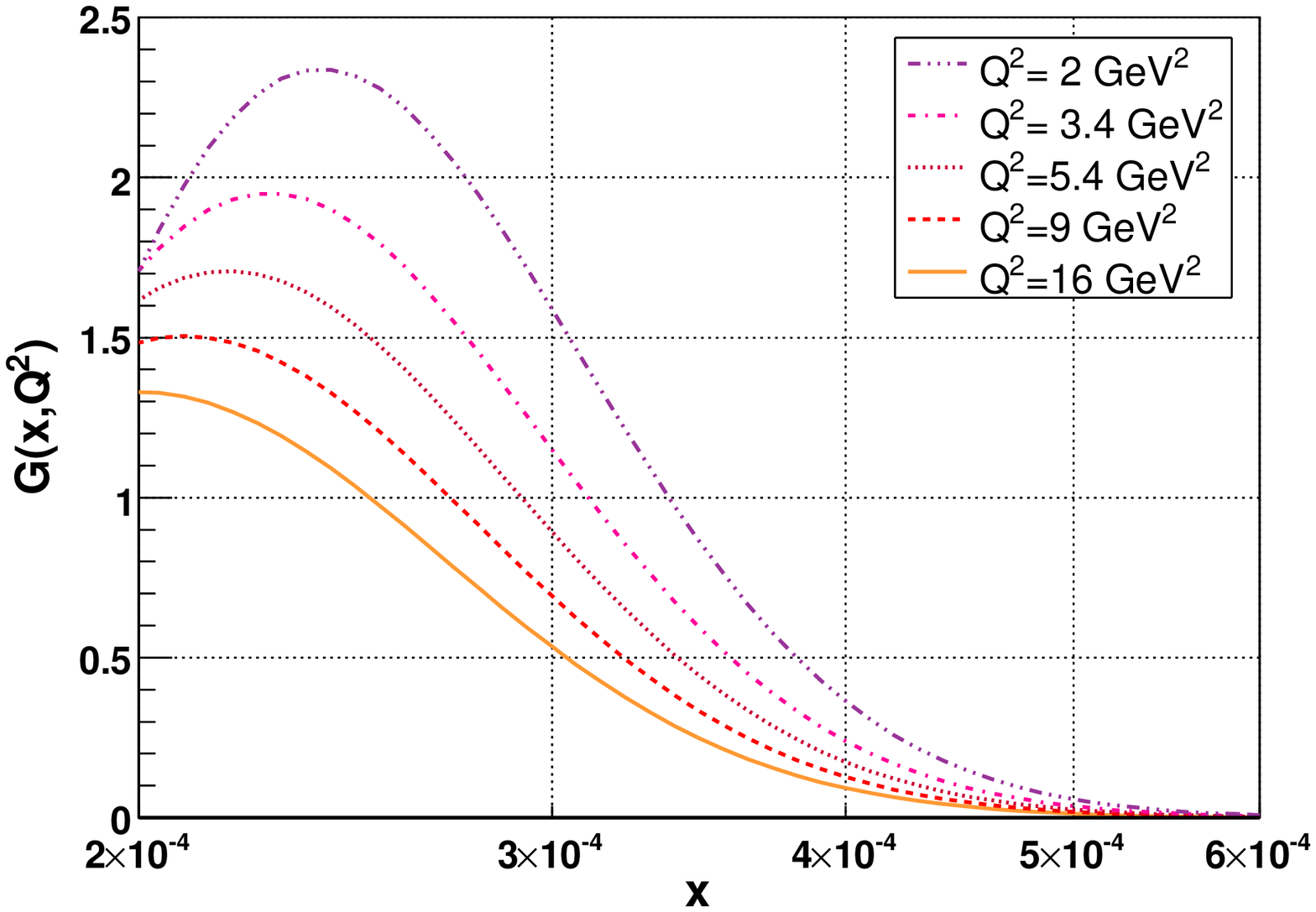, width=0.48\textwidth}
\epsfig{file=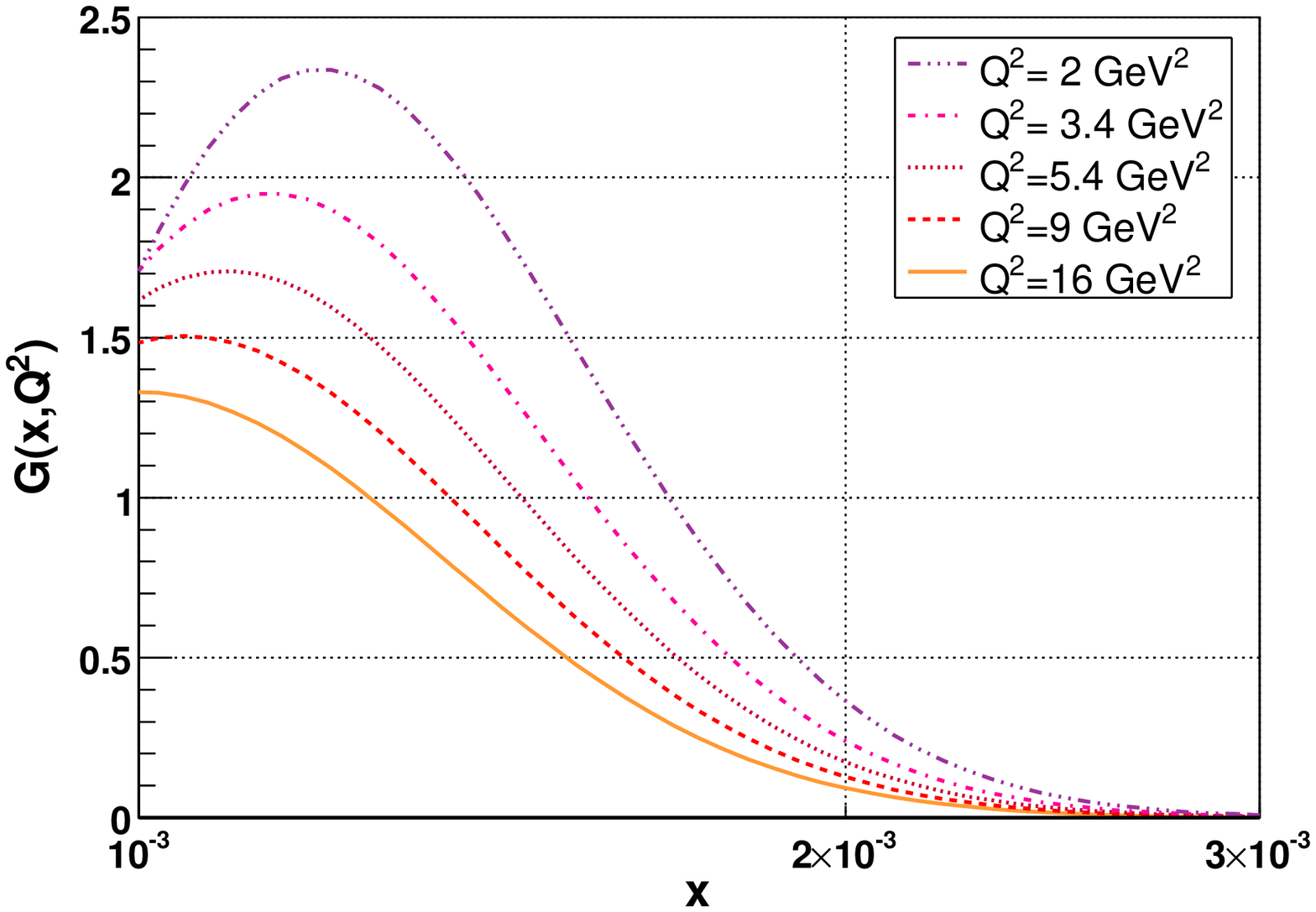, width=0.48\textwidth}
\caption{\small DGLAP backward evolution of a gaussian boundary condition $G\lp x ,Q^2_0=16 ~{\rm GeV}^2\rp$ centered at
$\bar x=2\cdot 10^{-4}$ (left) and 
$\bar x=  10^{-3}$ (right).}
\label{causalplot}
\end{figure}
% ----------------
The basic idea of this study is that if deviations from NLO DGLAP in
the data are hidden in a distortion of
parton distributions, they could be singled
out by determining undistorted PDF from data in  regions where
such effects are small~\cite{gelis}. 
We will do this in the following way: we determine PDFs 
using data at large $x$ and $Q^2$, where
 NLO DGLAP is likely to hold with high accuracy.
We then use NLO
DGLAP to evolve these PDFs down to the low $x$ and $Q^2$ region 
where deviations are expected to arise, and we compare our predictions
to the data in this region, which were not used in the PDF
determination. We then search for systematic deviations between data
and theory using a variety of statistical tools.

Possible deviations from fixed--order DGLAP could affect phenomenology
in two different ways. First, if they were indeed hidden by a
distortion of PDFs then this distortion could contaminate LHC
observables, which would thus be affected by a hitherto neglected
uncertainty. We will address this issue, which has already been raised
in the past with somewhat contradictory
conclusions~\cite{Martin:2003sk,Huston:2005jm} 
 in the last part of this paper. Second, such deviations  might provide
 evidence for effects which, if included systematically, could affect
 LHC observables in a non-negligible way. For instance, recent
 computation of small $x$ resummation corrections to various hard
 processes~\cite{Ball:2001pq,Ball:2007ra,Marzani:2008az,Marzani:2008uh,Diana:2009xv} shows that their effect at the LHC is expected to
 be of the same size or larger than NNLO corrections. Our  results may
 support the need for a systematic inclusion of  these effects.

% --------------------------------------------
\begin{figure}
\centering
\epsfig{file=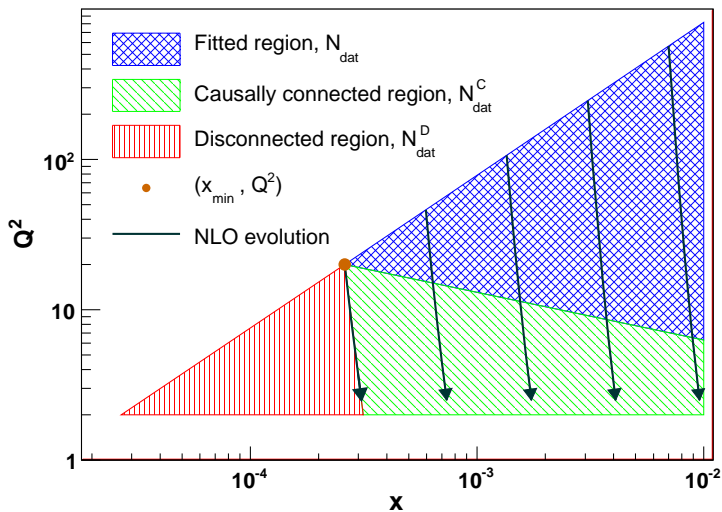,width=0.70\textwidth}
\caption{\small Causal structure of DGLAP evolution in the $(x,Q^2)$
  plane.
The arrow lines denote the trajectories followed by the maxima of the curves
in Fig.~\ref{causalplot}. The upper right (blue)  region contains the
data used to determined PDFs which are then evolved to the causally
connected (green) region  below it. 
No information can be obtained on
the (red) 'disconnected' region in the lower left corner. In
practice, the boundary between the connected and disconnected region
will be approximated by a vertical line with $x=x_{\rm min}$. The
number of data points in each region are listed in Table~\ref{tab:regions} below.
}\label{fig:kin-scheme}
\end{figure} 
% ---------------------------------------------

Because our basic strategy consists of comparing to data the results
of perturbative evolution, we must first discuss which kinematic 
regions are
connected by perturbative evolution in a causal way,
i.e., such that the results
of evolution to one region are affected by a change in the boundary
condition in the other region.
 The DGLAP evolution equation for the vector of PDFs $f(x,Q^2)$  has the form
\bq
\label{eq:dglap}
Q^2 \frac{d f\left(x,Q^2\right)}{d Q^2}  = \int_x^1\frac{dy}{y} P\left(\as(Q^2),\frac{x}{y}\right) f\left(y,Q^2\right) \ ,
\eq
where $P(\as, x)$ is a splitting function matrix. Because of the
convolution, the solution $f(\bar x,\bar Q^2)$ 
of Eq.~(\ref{eq:dglap})
at some 
point $z=(\bar x,\bar Q^2)$ only depends on the boundary 
condition $f(y, Q_0^2)$ in the 
range $y\in[\bar x,1]$.  Hence, a priori the past
causal cone of the point $(\bar x, \bar Q^2$) is given by the region
($x>\bar x$, $Q^2<\bar Q^2$). 

However,  the
bulk of the contribution to the convolution integral
Eq.~(\ref{eq:dglap}) comes from a small range in $x$, so that in
practice evolution mostly proceeds along trajectories that go along a
path from larger $(x_0,Q_0^2)$ to smaller $(\bar x,\bar Q^2)$. This
picture in fact becomes exact in the small $x$, large $Q^2$
limit, in which NLO DGLAP evolution is given by a wave equation which
may be studied using the method of
characteristics~\cite{Ball:1994du,Forte:1995vs}. 
In Fig.~\ref{causalplot} we show 
the  evolution of two gaussian boundary conditions centered 
around different $\bar x$ from $Q_{\rm in}^2=16$~GeV$^2$ down
 to $Q_{\rm fin}^2=2$~GeV$^2$, obtained solving the NLO DGLAP equation
 numerically with
the HOPPET~\cite{Salam:2008qg} package. Clearly, for the reasonably
short evolution
lengths shown, similar to those which we shall
consider in the sequel, the peaks remain  localized enough, and 
traverse
 approximately linear trajectories. These
trajectories are drawn as arrow lines in
Fig.~\ref{fig:kin-scheme}, and  turn out to be almost parallel
to the $y$--axis, consistent with the intuition that DGLAP evolution
is evolution in $Q^2$ at almost  fixed $x$. In conclusion, 
if we know parton
distributions for $x>\bar x$ at some scale $Q_0^2$  then DGLAP
evolution allows us to determine them for all $x>x_{\rm min}$ at any
other scale $Q^2$. However, for the evolution lengths displayed in
Fig.~\ref{causalplot}, $x_{\rm min}\approx \bar x$ to an accuracy of
a few percent. 

% ---------------------------
\begin{figure}
\centering
\epsfig{file=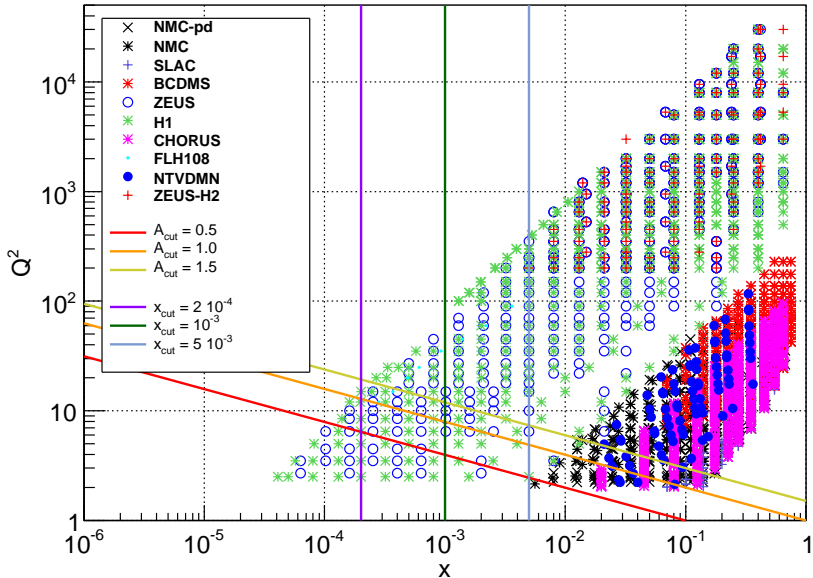, width=0.8\textwidth}
\caption{\small The NNPDF1.2 data set with
the various kinematic cuts applied in the present study.
}
\label{kincuts}
\end{figure}
% ---------------------------

We conclude that we can gain information on the low $x$, low $Q^2$ region
(where we expect possible deviations from NLO DGLAP) provided only we
have data at larger $Q^2$ in the same low $x$ region. However, as
$Q^2$ is raised, we
expect deviations to only arise at increasingly
smaller $x$. Indeed, we
tentatively expect deviations to appear when $Q^2<Q^2_{\rm cut}(x)$, with
\be
\label{eq:q2cutsat}
Q^2_{\rm cut}(x) \equiv  A_{\cut} x^{-\lambda} \ ,
\ee
where $\lambda\approx0.3$. This choice corresponds to the
kinematic region where the so--called geometric
scaling~\cite{Stasto:2000er} 
of structure
function data can no longer be understood as a consequence of
fixed-order DGLAP~\cite{Caola:2008xr} 
and could thus be evidence for effects beyond DGLAP. We will thus fit
data with $Q^2>Q^2_{\rm cut}(x)$  in order to gain information on the
low $Q^2$ region which is causally connected to it. These  regions are shown 
(for the choice $A_{\rm cut}=1.5$) in Fig.~\ref{fig:kin-scheme}.

We have thus performed a PDF determination based on the methodology
and dataset of the NNPDF1.2 parton set~\cite{Ball:2009mk}, but only
including data which pass the cut $Q^2>Q^2_{\rm cut}(x)$
Eq.~(\ref{eq:q2cutsat})  with various choices of $A_{\rm cut}$ and
$\lambda=0.3$. The data and various cuts are displayed in Fig.~\ref{kincuts};
the number of points in the fitted, causally connected, and
disconnected regions (defined as  in Fig.~\ref{fig:kin-scheme}) are
listed
in Table~\ref{tab:regions}. The fitted data include essentially all available  inclusive 
deep-inelastic scattering (DIS) data, as well as 
neutrino dimuon data, which are necessary in order to constrain the
strange PDF. The numbers given in Table~\ref{tab:regions}
for the data in the causally connected and
disconnected regions refer only to HERA data (not including $F_L$),
which will be used in the analysis below.

% ------------------------------
%
\begin{table}
\begin{center}
\begin{tabular}{|c|c|c|c|c|}
\hline
$A_{\rm cut}$ & $N_{\rm dat}$  & $N_{\rm dat}^{C}$& $N_{\rm dat}^{ D}$ & $(x_{\rm min}, Q^2 ~ [{\rm  GeV}^2])$ \\
\hline
no cuts & 3372 & 0 & 0 & $(4.1\cdot 10^{-5}, 2.5)$ \\
0.2 & 3363 & 4 & 5 & $(8\cdot 10^{-5}, 3.5)$ \\
0.3 & 3350 & 14 & 8 & $(10^{-4}, 6.5)$ \\
0.5 & 3333 & 25 & 15 & $(1.4\cdot 10^{-4}, 8.5)$ \\
0.7 & 3304 & 38 & 16 & $(1.6\cdot 10^{-4}, 12)$ \\
1.0 & 3228 & 44 & 19 & $(2.1\cdot 10^{-4}, 15)$ \\
1.2 & 3164 & 53 & 30 & $(2.4\cdot 10^{-4}, 15)$ \\
1.5 & 3084 & 59 & 38 & $(2.7\cdot 10^{-4}, 20)$ \\
\hline
\end{tabular}
\caption{\small 
Number of data points from Fig.~\ref{kincuts}
in the regions of the $(x,Q^2)$ plane defined according to
Fig.~\ref{fig:kin-scheme}
with the
cut Eq.~(\ref{eq:q2cutsat}).
 The columns
show, from left to right:
the value of $A_{\rm cut}$ Eq.~(\ref{eq:q2cutsat}) used
in the cut; the 
total number of points $N_{\rm dat}$
which pass the cut; the number of points $N_{\rm dat}^C$ in the
causally connected region; 
the number of points $N_{\rm dat}^D$ in the disconnected
region; the minimum value of  $(x_{\rm min}, Q^2)$ for the data region
(coordinates of the point denoted by a dark (brown) dot in
Fig.~\ref{fig:kin-scheme}). The total number of points
 $N_{\rm dat}$ refers to
all experiments shown in Fig.~\ref{kincuts}, while the number of points in
the connected and disconnected regions refer to HERA data only
(not including $F_L$ data). 
\label{tab:regions}}
\end{center}
\end{table}
%
% ------------------------------

The NNPDF
methodology~\cite{DelDebbio:2007ee,Ball:2008by}
is especially suited to this analysis because it provides a 
determination  of PDFs and their uncertainty which is  independent of the
choice of data set, and which has been shown in benchmark
studies~\cite{Dittmar:2009ii} to behave in a statistically consistent
way when data are added or removed to the fit. Also, because of the use of a
Monte Carlo approach, the NNPDF methodology is easily amenable 
to the use of standard
statistical analysis tools as we shall see below. This methodology has
been applied successfully to determination of
parton
distributions~\cite{DelDebbio:2007ee,Ball:2008by,Ball:2009mk,Rojo:2008ke,Guffanti:2009xk}, unpolarized~\cite{Forte:2002fg,DelDebbio:2004qj}  and
polarized~\cite{DelDebbio:2009sq}  structure
functions, 
QCD spectral 
functions~\cite{Rojo:2004iq} and atmospheric neutrino
fluxes~\cite{GonzalezGarcia:2006ay}.

The main
drawback of current NNPDF parton fits is the way heavy quarks are
treated, namely, the fact the
so--called zero--mass variable--flavour number scheme is used. This
means that terms which are suppressed by powers of heavy quark mass
over the large scale of the process are neglected, which is
 a poor approximation
close to the threshold for heavy quark production. This may be a
problem for our analysis because most of the data we are interested in
is close to the charm threshold. This issue will have to be addressed
when analyzing our results.

%----------------------------------------------------------------
\begin{figure}
\epsfig{file=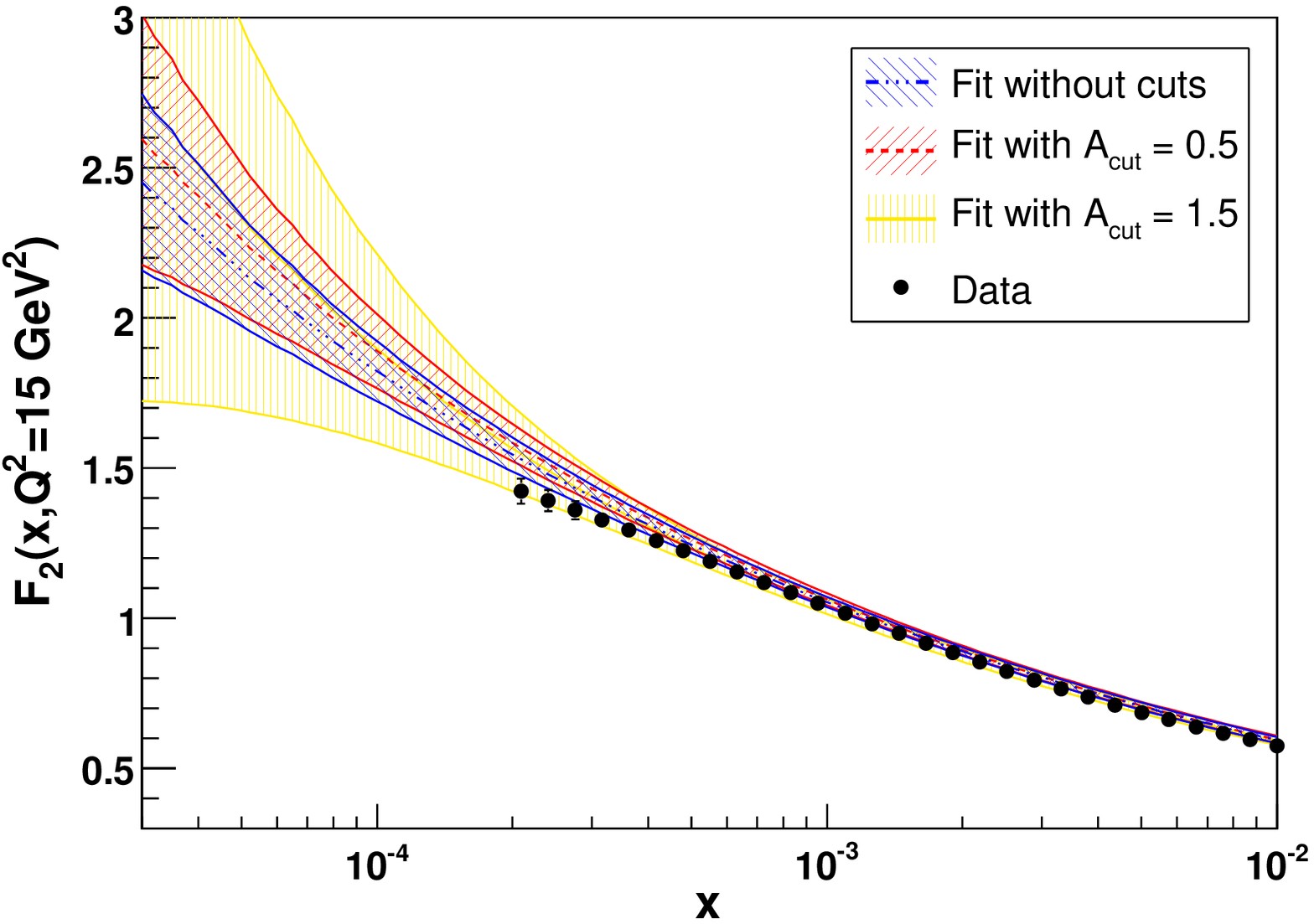, width=0.5\textwidth}
\epsfig{file=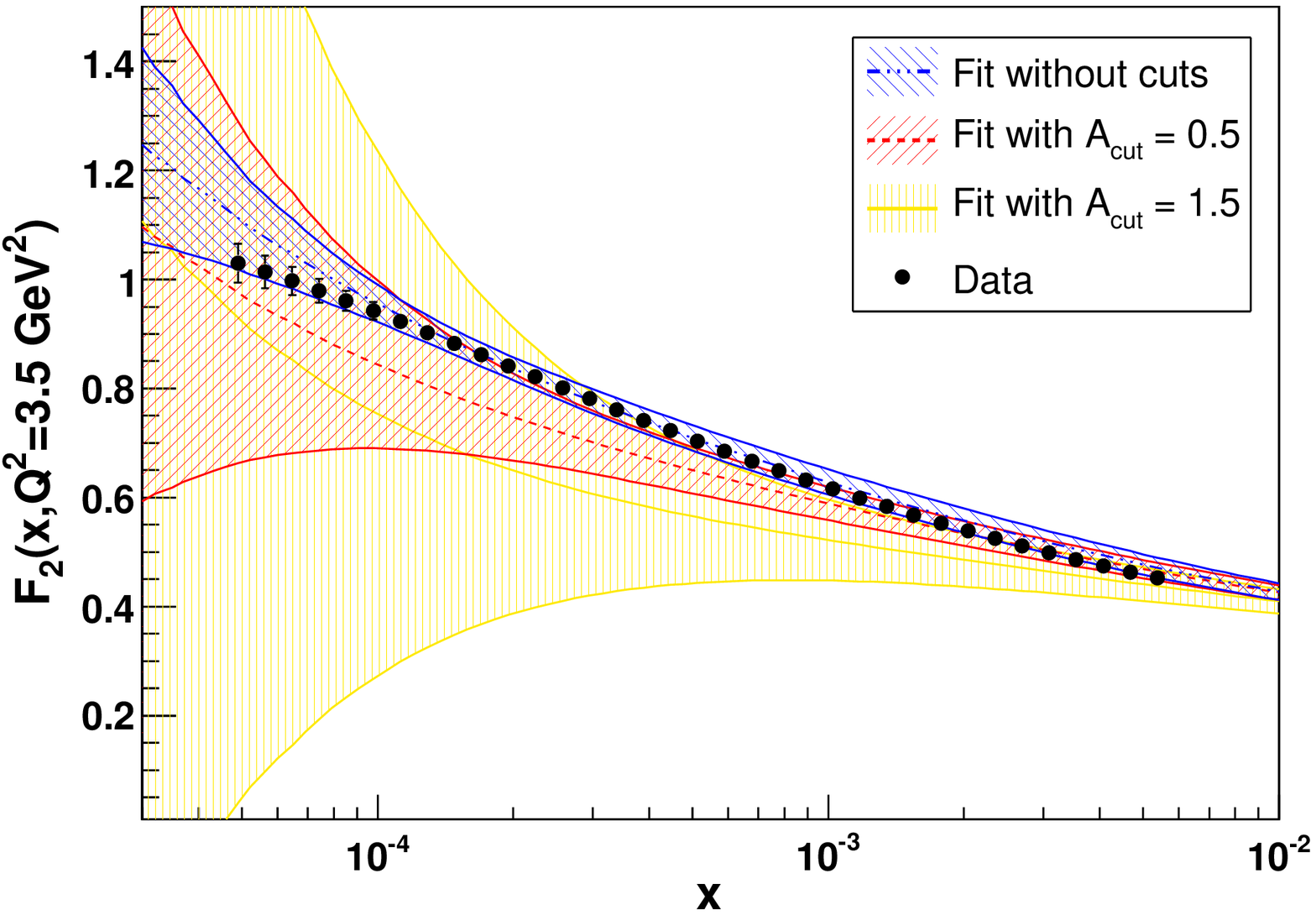, width=0.5\textwidth}
\caption{\small The proton structure function $F_2(x,Q^2)$ at small $x$, 
computed from PDFs obtained from fits with different
values of $A_{\rm cut}$, for $Q^2=$15 GeV$^2$ (left) and
$Q^2=$3.5 GeV$^2$ (right), compared to the data. Wider uncertainty bands
correspond to more restrictive cuts.}\label{f2_plot}
\end{figure}
%-----------------------------------------------------------------

The results of our fits with various
cuts are compared to the standard
NNPDF1.2 fit and to experimental data in Fig.~\ref{f2_plot}.
We show the structure function $F_2(x,Q^2)$  in
the small $x$ region, both at a scale in the data region (left) and
at a low scale in the region which is causally connected to
it (right), with no cut (standard NNPDF1.2 fit~\cite{Ball:2009mk}) and
with the lowest and highest of the cuts of Fig.~\ref{kincuts}. Instead
of showing directly the data used in the fit, we display the very
precise interpolation of the data of Ref.~\cite{Del Debbio:2004qj},
which is more accurate than any individual data point because it
combines all data in a  way which does not depend on theory or model
assumptions. 

It is clear that 
at the high 
scale $Q^2=15$~GeV$^2$ there is no 
significant difference in the data region
between the three different predictions from
the fit without cuts, the one with intermediate cut and
the one with the maximum cut. The only difference is the growth
of the PDF uncertainty in the extrapolation region, which is
statistically expected due to the missing experimental information
removed by the cuts.
However, at low  $Q^2=3.5$~GeV$^2$, besides showing an increase of
uncertainty, the prediction obtained by backward evolution of the data
above the cut
 exhibits a systematic downwards trend: 
it always lies below the
HERA data. As we increase the value of $A_{\cut}$, this trend becomes more and
more evident, and it turns on smoothly as we move from $Q^2=15$~GeV$^2$
to $Q^2=3.5$~GeV$^2$. This is to be contrasted to the uncut NNPDF1.2 fit,
which always sits on top of the data.
 It would thus seem that  the backward NLO DGLAP evolution of the
 high--scale data is too strong: it overestimates the actual amount of
 evolution seen in the data themselves.

\begin{figure}
\epsfig{file=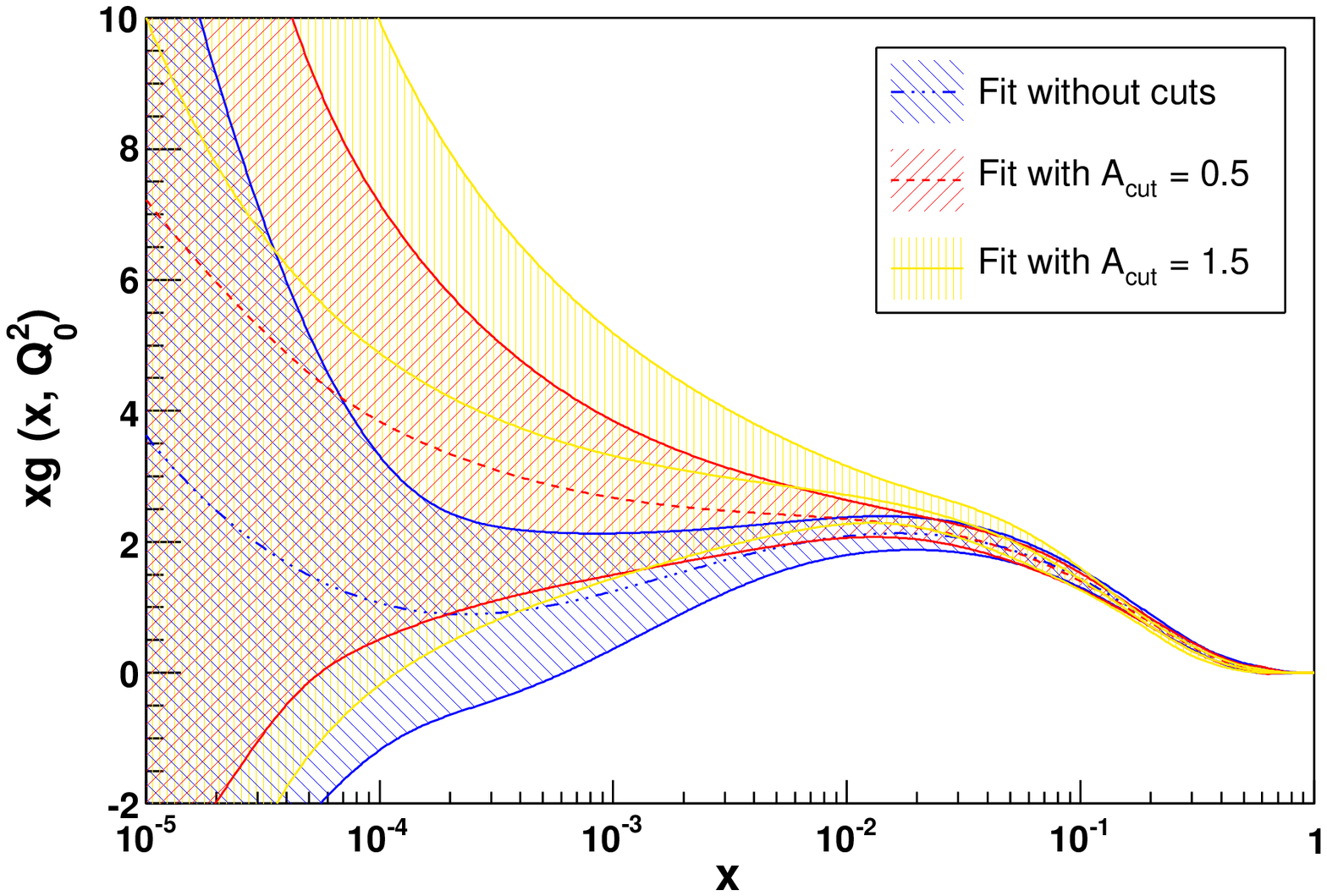,width=0.5\textwidth}
\epsfig{file=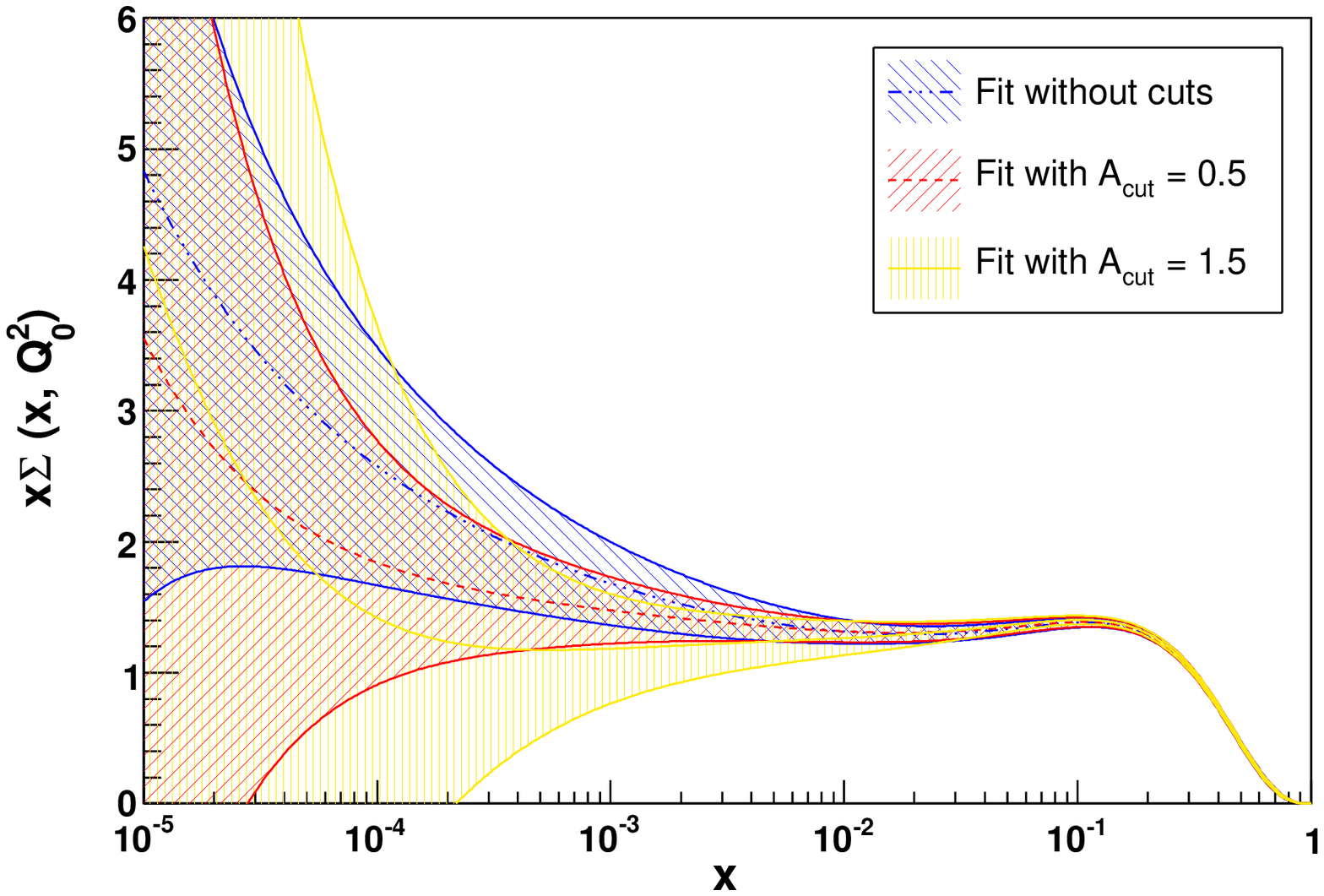,width=0.5\textwidth}\\
\epsfig{file=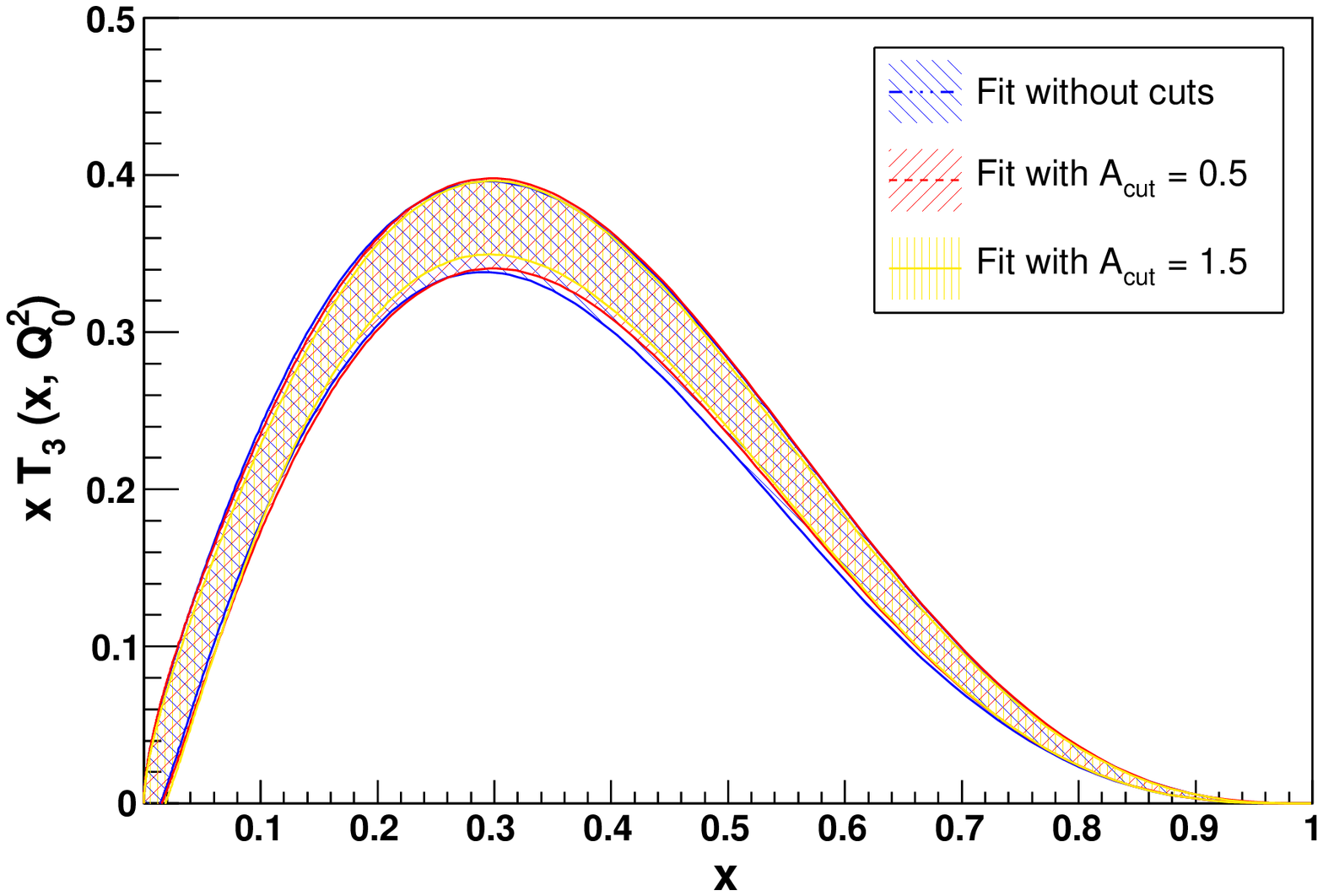,width=0.5\textwidth}
\epsfig{file=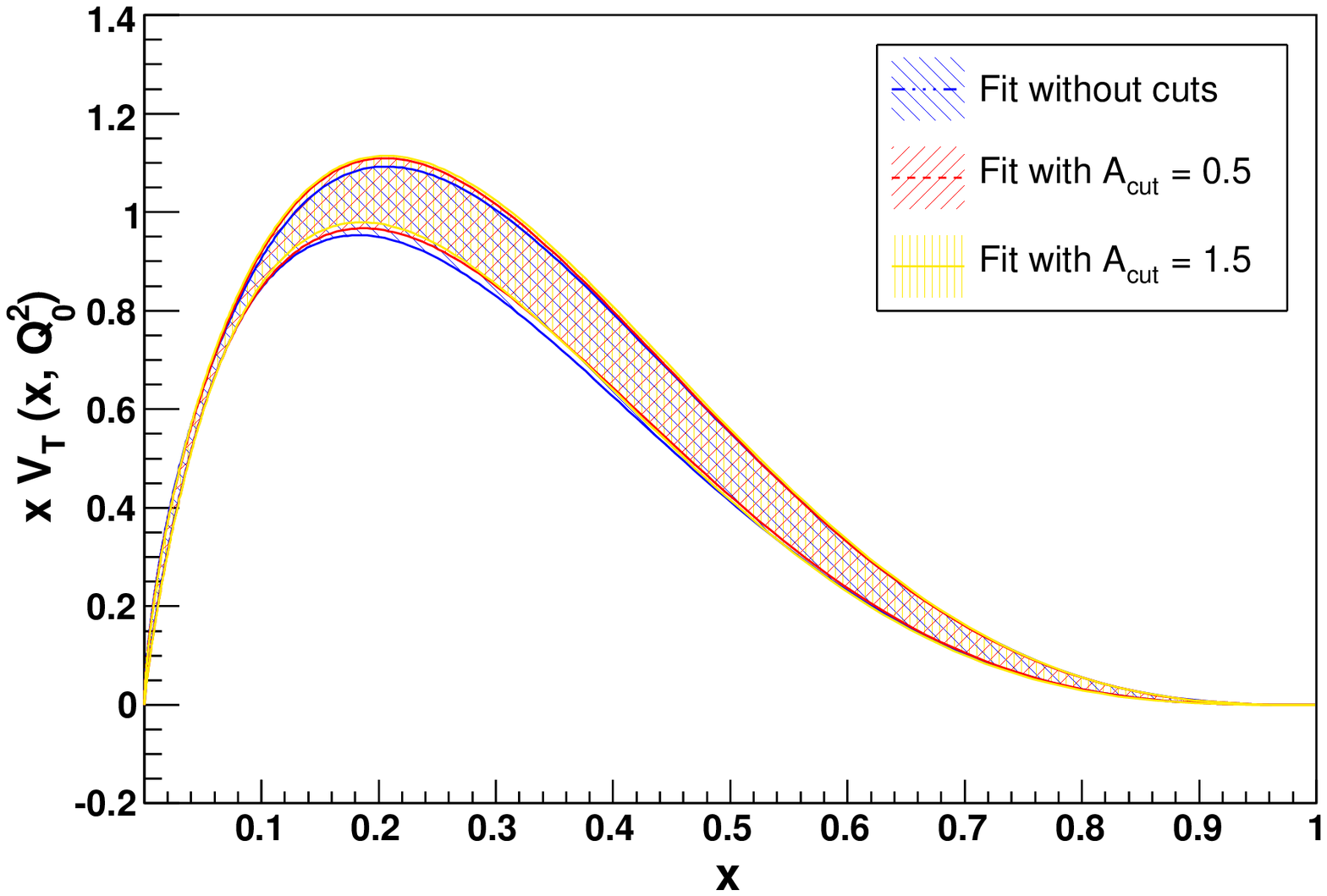,width=0.5\textwidth}\\
\caption{\small The gluon, singlet, triplet and total valence 
PDFs with different cuts at the
scale $Q_0^2=2$~GeV$^2$. The gluon and singlet are plotted on a log
scale and the valence and triplet on a liner scale  to emphasize
respectively the small $x$ and large $x$ regions.}
\label{pdfq0}
\end{figure}

Before trying to assess more quantitatively the size of this effect,
let us first examine  how the   cuts
affect the individual PDFs. In Fig.~\ref{pdfq0} we show the singlet and
gluon PDFs, which are largest at small $x$; as well as the valence and
triplet which dominate at large $x$,
for the uncut fit and for two different kinematical cuts.
We observe that  at
small $x$ the cut produces
a sizable increase  in PDF uncertainties  and a change in
central values which seems to follow a systematic trend as the cut is
moved.  However, PDFs 
are  consistent with each other  at the
one sigma level, which implies that predictions  for physical observables
obtained from any of these PDFs will also 
be compatible at this level, as we shall see explicitly below.
On the other hand, at large $x$ the PDFs are essentially
unaffected by the cut. This shows that the effect of the cut is indeed
only on the region affected by it, displayed  in Fig.~\ref{fig:kin-scheme}.

A quantitative estimate of possible deviations can be obtained 
by defining  the statistical distance between a
data point $F_{\data,\,i}$ with uncertainty $\sigma_{\data,i}$
and the associated theoretical prediction $F_{{\rm th},\,i}$ with uncertainty 
 $\sigma_{{\rm th},i}$
\bq\label{dist}
d_{i}^{\rm stat}\equiv\frac{F_{\data,\,i}-F_{{\rm th},\,i}}{\sqrt{\sigma_{\data,\,i}^2+\sigma_{{\rm th},\,i}^2}} \  .
\eq
where $\sigma_{{\rm th},\,i}$ stands for the PDF uncertainty.
In order to determine the absolute scale of these deviations from
NLO DGLAP,  we also define
the relative distance 
\bq\label{percdist}
d_{i}^{\rm rel}\equiv
\frac{F_{\data,i}-F_{{\rm th},i}}{\left(F_{\data,i}+F_{{\rm th},i}\right)/2}
\ .
\eq
Whereas $d_{i}^{\rm rel}$ measures the  absolute size of
the deviation,
$d_{i}^{\rm stat}$  measures its statistical significance in unit of
the standard deviation: for
instance $d_{i}^{\rm stat}=1$ means that the deviation is at the
one--sigma level.
% ----------------------------
% -----------------------------
% --------------------------------------------------
\begin{figure}
\epsfig{file=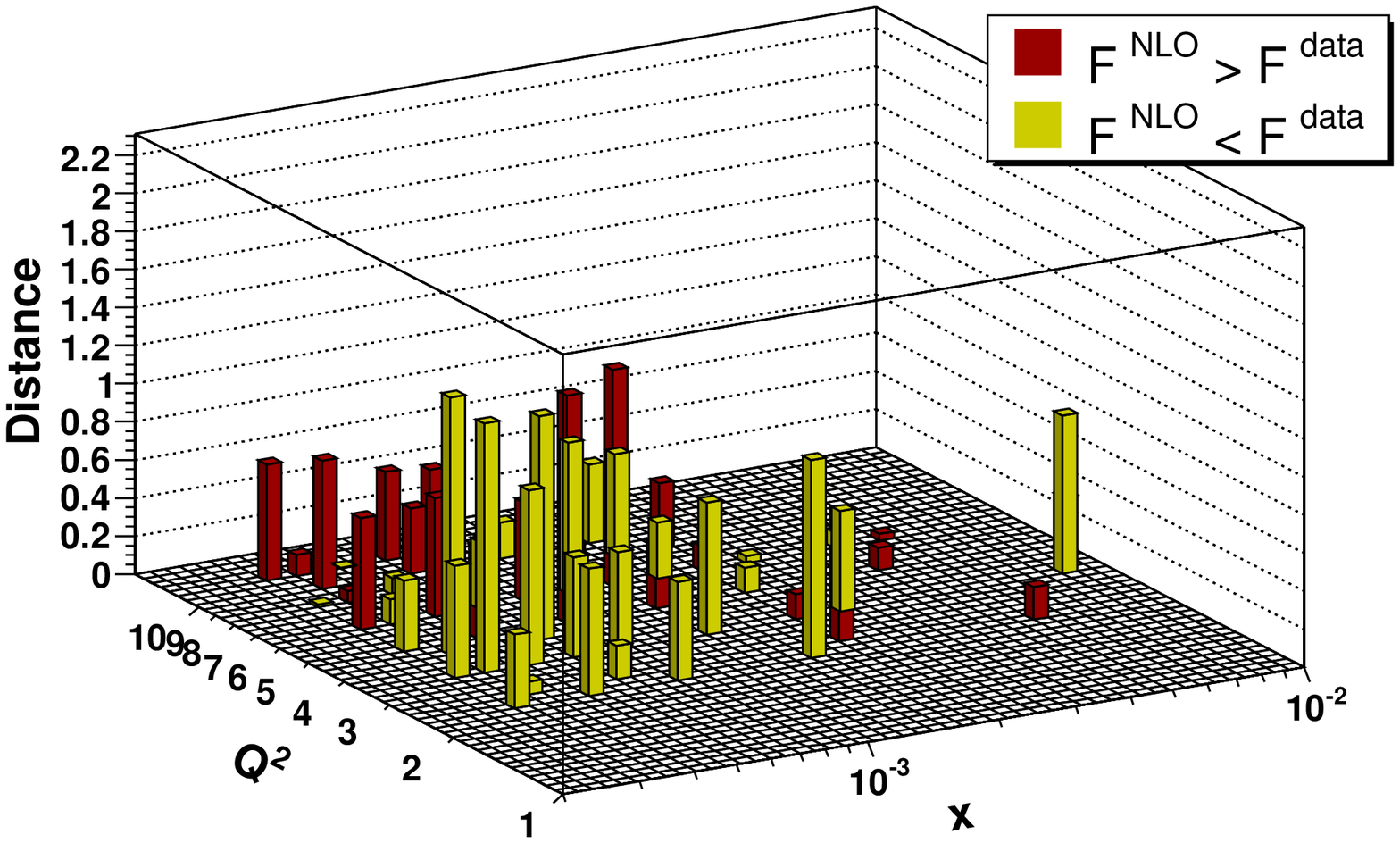,width=0.48\textwidth}
\epsfig{file=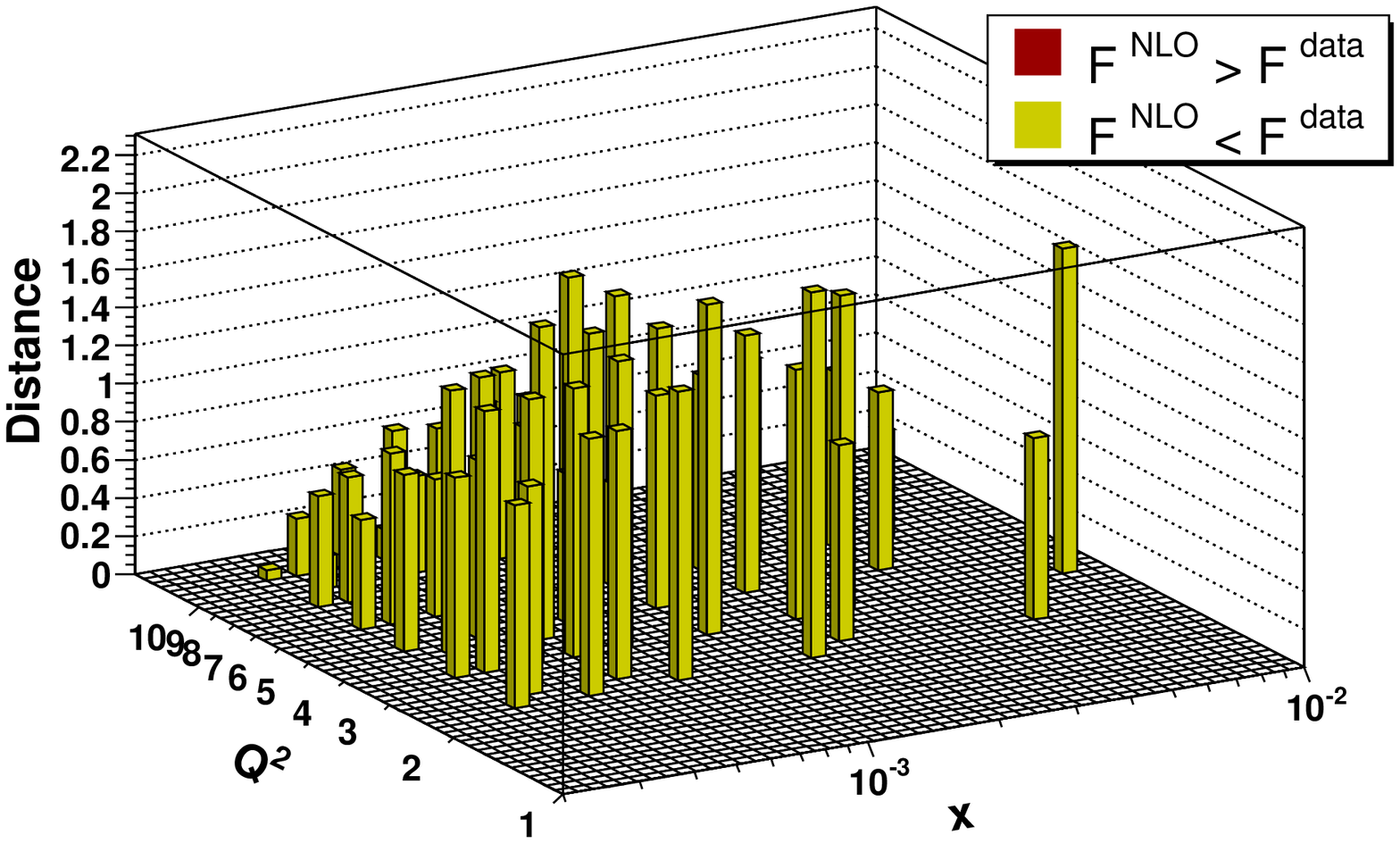,width=0.48\textwidth}
\caption{\small Statistical distance, Eq.~(\ref{dist}), between the data
  in the region excluded from the fit but causally connected to it,
  and the prediction obtained using NLO DGLAP evolution of the data
  included in the fit, with $A_{\cut}=1.5$.  
The distance is computed both when the cut is not applied (left) and when
it is applied
(right). Dark (red) bars denote negative distances (fit above the
data) while light (yellow) bars denote positive distances  (data above
the fit). The average values of these
 distances are listed in Table~\ref{stat_dist_table}. 
}\label{distance} 
\end{figure} 
% ----------------------------------------------------

In Fig.~\ref{distance} (right plot) we show the statistical 
distance Eq.~(\ref{dist}) between
the prediction obtained from backward NLO DGLAP evolution of the fit
to data above the cut to its
causally connected region, and data in this region, for the
$A_{\rm cut}=1.5$ case.
As a control sample,  we also compute and show (left plot) the
distance 
when the data in the cut region are not excluded from the
fit. Distances are computed using HERA data only (not including $F_L$
data) because of their greater consistency (non--HERA data in these
regions are extremely scarce anyway).
Comparing the two plots, it is clear that when all data are included
in the fit the sign of the distances are distributed randomly (the fit
is equally likely to be above or below the data), and
typically
$d^{\rm stat}\lsim 1$. However, when the data are excluded from the plot,
the fit tends to systematically undershoot the data (almost all
distances are positive), while their size is systematically somewhat
larger, 
$d^{\rm stat}\gsim 1$. Also, the size of the distance tends to
increase somewhat if the data are further away from the cut.

The  same conclusion is obtained by inspecting
the values of the distance, which are tabulated in 
Table~\ref{stat_dist_table} with various cuts, and in the various
regions which are excluded by them. Here we also show, for each value
of the distance, the statistical uncertainty with which it is
determined, as estimated from the variance of the results obtained
from the  set of PDF replicas.
Namely, we find that if no cut is applied, the
distances are small and fluctuate randomly, by an amount which is
comparable to the statistical uncertainty with which the distance is
computed. 
Once a cut is applied, the
distances become all positive, they differ from zero by a
statistically significant amount, and they show a tendency to increase
when the data are further away from the cut. This confirms the
qualitative features seen in Fig.~\ref{distance}: there is evidence for
a statistically significant deviation of the data from the NLO DGLAP
prediction. The evidence is only at the one--sigma level
(i.e. $d^{\rm stat}\sim1$) , but it is
systematic, and its significance tends to increase when the amount of
evolution required to get to the given point or region is larger,
despite the fact that points with longer evolution lengths are those at
the lowest $x$ and $Q^2$ and thus affected by the largest uncertainties.

% --------------------------------
\begin{table}
\centering
\begin{tabular}{|c|c|c|c|c|}
\hline
$A_{\cut}$ for the fit & \multicolumn{4}{|c|}{
$\la d^{\rm stat}\ra$} \\
\cline{2-5}
 & $A_{\cut}<0.5$ & $0.5<A_{\cut}<1.0$ & $1.0<A_{\cut}< 1.5 $ & $A_{\cut}< 1.5$ \\
\hline
no cuts & 0.6 $\pm$ 0.5 & -0.1 $\pm$ 0.5 & -0.1 $\pm$ 0.3 & 0.06 $\pm$ 0.6\\
0.5 & 1.4 $\pm$ 0.4 & --- & --- & 1.4 $\pm$ 0.4\\
1.0 & 1.2 $\pm$ 0.2 & 0.7 $\pm$ 0.3 & --- &  0.9 $\pm$ 0.4 \\
1.5 & 1.4 $\pm$ 0.3 & 0.9 $\pm$ 0.4 & 0.6 $\pm$ 0.4 & 1.0 $\pm$ 0.5 \\
\hline
\end{tabular}
\caption{\small 
Average statistical distance, Eq.~(\ref{dist}),
 between the data
  in the region excluded from the fit but causally connected to it,
  and the prediction obtained using NLO DGLAP evolution of the data
  included in the fit. The uncertainty given is the statistical.
Each row corresponds to a different cut and
  thus a different set of data; the case $A_{\rm cut}=1.5$ corresponds
  to the distances shown in
  Fig.~\ref{distance}. Each column gives the average over
  the subset of data in the pertinent row which would also pass
  various less restrictive cuts, i.e. from left to right data in
   regions which are increasingly close to the cut corresponding
  to the given row. In each case, we give the average and
standard deviation of the distance for all points included in the
corresponding region. All HERA data (not including $F_L$ data)
shown in Fig.~\ref{kincuts} are used; total numbers of data points for
each cut are listed
in Tab.~\ref{tab:regions}. 
 }\label{stat_dist_table}
\end{table}
% ---------------------------------------

\begin{figure}
\begin{center}
\epsfig{file=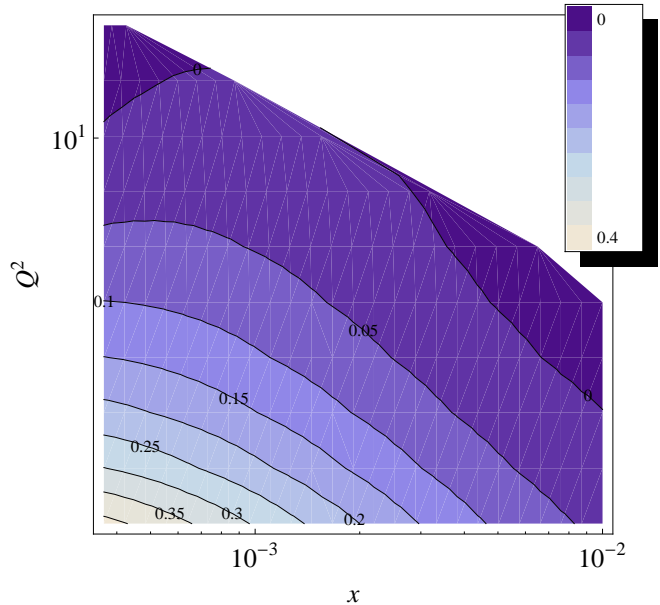,width=0.6\textwidth}
\end{center}
\caption{\small Contour plot of the relative distances. Eq.~(\ref{percdist}),
with  the most restrictive  cut $A_{\rm cut}=1.5$. Distances are
computed between the NLO DGLAP and the  parametrization of the data
Ref.~\cite{DelDebbio:2004qj}.}  
\label{distanceperc}
\end{figure}

%-----------------
\begin{table}
\centering
\begin{tabular}{|c|c|c|c|c|}
\hline
$A_{\cut}$ for the fit & \multicolumn{4}{|c|}{$\la d^{\rm rel}\ra$} \\
\cline{2-5}
 & $A_{\cut}<0.5$ & $0.5<A_{\cut}<1.0$ & $1.0<A_{\cut}< 1.5 $ & $A_{\cut}< 1.5 $  \\
\hline
no cuts & 0.03 $\pm$ 0.03 & -0.004 $\pm$ 0.02 & -0.006 $\pm$ 0.01 & 0.01 $\pm$ 0.03\\
% 0.2 & 0.0005 $\pm$ 0.02 & / & / & 0.0005 $\pm$ 0.02 \\
%0.3 & 0.05 $\pm$ 0.02 & / & / & 0.05 $\pm$ 0.02 \\
0.5 & 0.13 $\pm$ 0.05 & --- & --- & 0.13 $\pm$ 0.05 \\
%0.7 & 0.15 $\pm$ 0.05 & 0.04 $\pm$ 0.02 & / & 0.12 $\pm$ 0.07 \\
1.0 & 0.22 $\pm$ 0.06 & 0.06 $\pm$ 0.03 & --- & 0.13 $\pm$ 0.09 \\
%1.2 & 0.17 $\pm$ 0.05 & 0.05 $\pm$ 0.03 & 0.03 $\pm$ 0.02 & 0.09 $\pm$ 0.07\\
1.5 & 0.27 $\pm$ 0.07 & 0.08 $\pm$ 0.03 & 0.03 $\pm$ 0.02 & 0.12 $\pm$ 0.11\\
\hline
\end{tabular}
\caption{\small Same as Table~\ref{stat_dist_table} but for the relative distance Eq.~(\ref{percdist}).}\label{perc_dist_table}
\end{table}
% --------
We can study the absolute
size of the effect by repeating the same analysis, but now for the relative
distance Eq.~(\ref{percdist}). Results are collected in
Table~\ref{perc_dist_table}: it is clear that the deviation increases
in size as the region of the comparison is further away from the
fitted region, 
as one would expect of a deviation driven by the evolution length.
A contour plot of the distances, Fig.~\ref{distanceperc}, clearly
shows the increase of the distance with the evolution length.

% -----------------------------
% ----------------------------

\begin{figure}
\centering
\epsfig{file=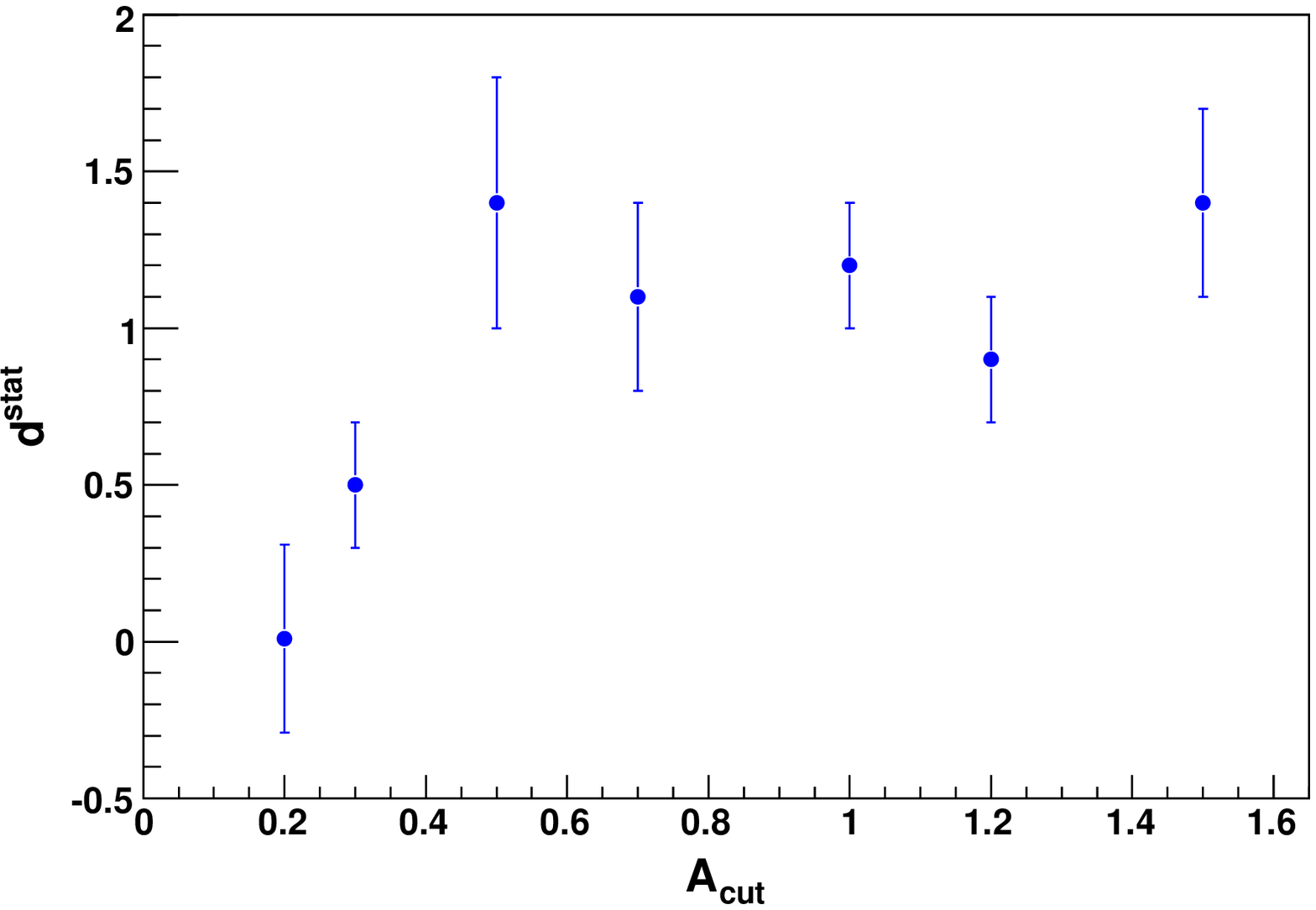,width=0.45\textwidth}
\epsfig{file=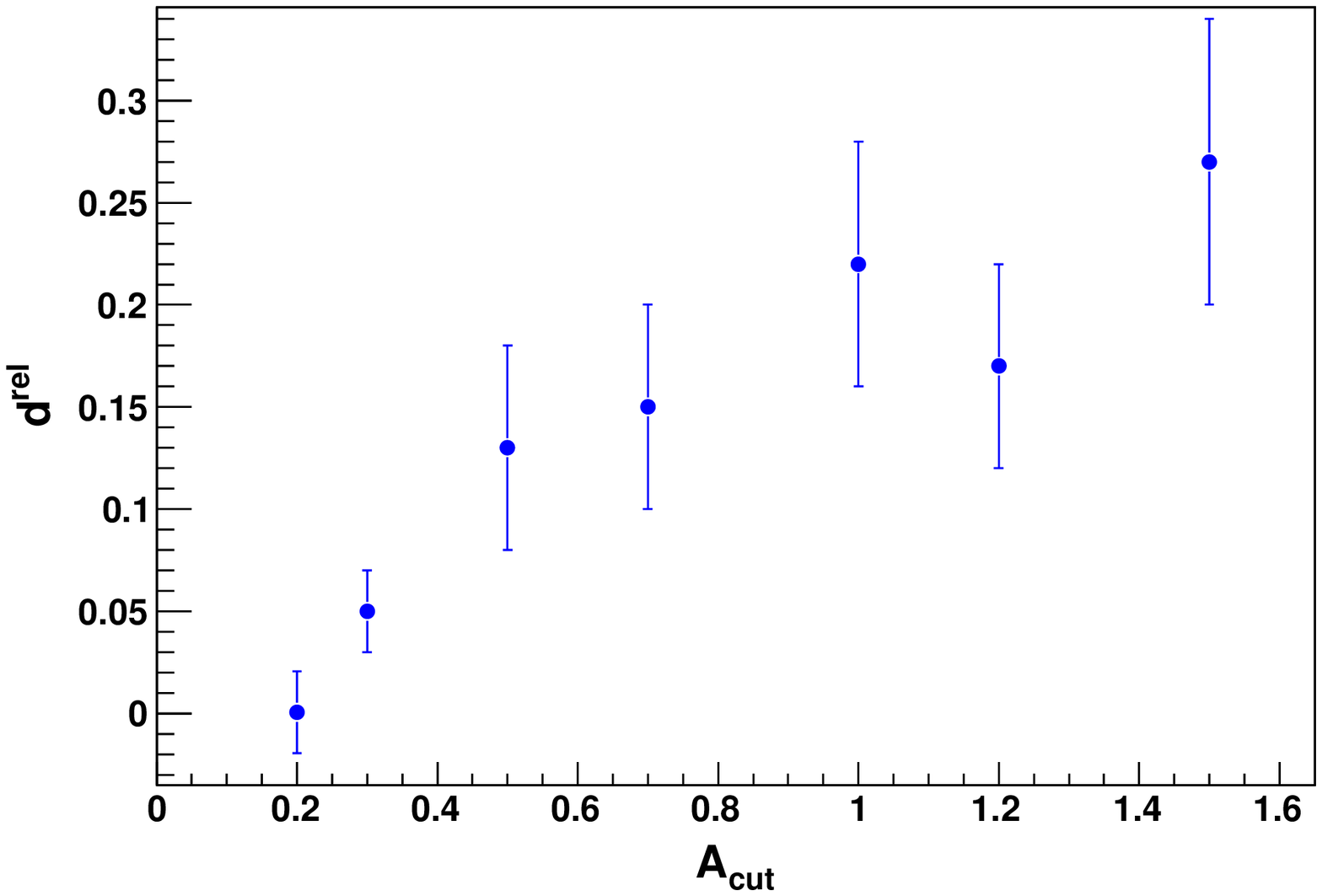,width=0.45\textwidth}
\caption{\small Statistical distances Eq.~(\ref{dist}) (left) and
  relative distances Eq.~(\ref{percdist})
(right) for fits with different $A_{\cut}$
as a function of $A_{\cut}$, for all the choices of cuts of
Table~\ref{tab:regions}. The distances shown are all computed for the
subset of points of each fit which also satisfies the cut
$A_{\cut}<0.5$, i.e. they correspond to the first column of
Tables~\ref{stat_dist_table}-\ref{perc_dist_table}. 
}\label{da_plot}
\end{figure}

The behaviour of the statistical and relative distances as a function
of $A_{\rm cut}$ is summarized in
Fig.~\ref{da_plot}, where we show the distances in the
first column of Tables~\ref{stat_dist_table}-\ref{perc_dist_table},
including also the further intermediate values of $A_{\cut}$ shown in
Table~\ref{tab:regions}. These values correspond in each case to the data
in the causally connected region which are further away from the
cut. The increase of the relative distance (right), and the general trend of
an increase of its significance (left) can both be seen clearly.

As a final piece of evidence, we also show in Table~\ref{pdf-sat}   
the $\chi^2$ for the HERA data 
points of Table~\ref{tab:regions} excluded by various cuts, in the 
causally connected region for each cut, and compare them to the values
obtained in the absence of a cut. For reference, we also
 show the $\chi^2$ of the
non-HERA data points, which are at large $x$ and only marginally
affected by the cuts.
In each case, the quality of the fit of the HERA data
when they are not fitted is significantly worse, and it gets worse
when the cut is more restrictive, i.e. more data require more
evolution. In comparison, the quality of the fit of the large $x$
(non-HERA) data is unaffected up to small fluctuations which do not
show any systematic trend as the cut is raised.
 Note that both in the NNPDF
fits~\cite{Ball:2008by,Ball:2009mk} 
and in benchmarks~\cite{Dittmar:2009ii} it has been explicitly checked 
that when groups of data (in individual regions, or from particular
experiments) are
removed from the NNPDF fit, the quality of the NNPDF
fit to those data does not deteriorate, because uncertainty bands
always widen in such a way that the statistical compatibility of the
fit with the excluded data
is unchanged. Hence, we must conclude that in this specific kinematic
region  there is a
deviation from the fit which is statistically
significant and whose origin is not statistical. 

% ----------------------------
% -----------------------------------------------------------------------

We conclude therefore that we have a statistically significant
indication that data at low $x$ and $Q^2$ deviate from the prediction
of NLO DGLAP evolution. It looks very unlikely that this deviation is
a byproduct of specific features of the fit. Specifically, we have
shown (see Figs.~\ref{causalplot}-\ref{fig:kin-scheme}) that evolution
from the boundary into the region affected by the cuts is essentially
local, so that what one observes is indeed a deviation from the
predicted evolution rate, while global aspects which are in
principle present due to the integro--differential nature of DGLAP
evolution do not come into play. Furthermore, we have explictly seen
(see Fig.~\ref{pdfq0} and Table.~\ref{pdf-sat}) that the behaviour of the
fit in the large $x$ region is completely unaffected by the cuts,
which makes it very unlikely that the discrepancy could be removed by
modifying some features of the global fit such as e.g. wild
modification of large $x$ PDFs. Also, the effect appears to be
statistically significant, while the general features of the fit such
as widening of error bands and the size of fluctuations are 
completely  consistent with statistical expectations.

One must therefore ascribe the effect to some systematic theoretical source, and
ask what it may be.
One source is readily identified, namely, heavy quark mass effects. 
Indeed, as already mentioned, the zero--mass variable flavour number
scheme used by NNPDF is not accurate in the charm threshold region,
which largely overlaps with the region under consideration here.
These effects go in the same direction as the deviation found here:
the charm quark mass suppresses perturbative evolution driven by charm
radiation and its mixing with gluons~\cite{Nadolsky:2009ge}. However,
the size of these effects was recently assessed quantitatively in 
Ref.~\cite{Tung:2006tb}, where it was found that it is never
above  $15\%$ for $F_2$, while we find 
deviations as large as $35\%$ at the smallest $x$ and $Q^2$ values. We
conclude that even
though heavy quarks could be partly responsible for the effect  that
we have seen, they are unlikely to be the only explanation.

We therefore look at other possible reasons for deviation from NLO
DGLAP predictions. Whereas a detailed quantitative study is 
beyond the scope of this work, we briefly examine whether some known
potential sources of deviations from the NLO DGLAP prediction are
qualitatively 
compatible with our findings. Such deviations could be due to higher
order perturbative effects (such as usually estimated by varying the
renormalization and factorization scales), or to higher twists.
The simplest option for 
higher order perturbative corrections is fixed--order NNLO terms, which are
obviously more important at low scale, and are  known to grow at
small $x$. However, NNLO corrections at small $x$ are known to lead to
a stronger scale dependence of $F_2(x,Q^2)$ in comparison to the NLO
prediction. This can be seen both in global NNLO
fits~\cite{Alekhin:2001ih,Martin:2009iq} and from a study of anomalous
dimensions and $K$--factors obtained from
them~\cite{resum,Rojo:2009us}. Therefore, even though the precise size
of these effects can only be estimated within a full NNLO fit, they
can only make things worse.

% -----------------------------------------------------------------------
\begin{table}
\centering
\begin{tabular}{|c|c|c||c|}
\hline
$A_{cut}$ & \multicolumn{2}{|c||}{ HERA data} & Non-HERA data \\
\cline{2-4}
&   $\chi^2_\text{ without cuts}$ & $\chi^2_\text{ with cut}$ & $\chi^2$ \\
\hline
no cuts & --- & --- & 3020/2098 = 1.44\\
0.5 & 19.68/25 = 0.79 & 106.22/25 = 4.25 & 2973/2098 = 1.42\\
1.0 & 54.41/44 = 1.24 & 138.24/44 = 3.14 & 2943/2098 = 1.40\\
1.5 & 62.31/59 = 1.06 & 860.65/59 = 14.6 & 2955/2098 = 1.41\\ \hline
\end{tabular}

\caption{\small Left column: the $\chi^2$ for all the points
  excluded by a cut, 
 in the
  region causally connected to the cut, as described by the
third column in Table~\ref{tab:regions}, 
compared to the $\chi^2$ for the same points in the fit without
cuts. Right column: the $\chi^2$ for the non-HERA data.}
\label{pdf-sat}
\end{table}
% ---------------------------------------------------------------------

However, it is well--known that perturbation theory becomes unstable
at small $x$ and must be resummed. The all--order resummation of small
$x$ corrections to perturbative evolution, properly matched to fixed--order
DGLAP evolution~\cite{resum}, in the
HERA region leads to an effect that is qualitatively opposite to that
of NNLO corrections, namely to a weaker scale dependence in comparison
to NLO, as one can again see from a study of anomalous
dimensions and $K$--factors obtained from
them~\cite{resum,Dittmar:2009ii,Rojo:2009us}. It follows that small
$x$ resummation  could explain the deviation that we observe.

Perturbative evolution at small $x$ is expected to eventually  be
corrected by non--linear terms, related to high parton densities, 
which should restore unitarity in the
very high energy limit~\cite{GolecBiernat:2008nq}, and which in a
perturbative framework appear as higher twist corrections.
The precise signature of these effects is not easy to assess, because 
it is nontrivial to match these effects to DGLAP evolution, hence
their predictions are usually given close to the asymptotic
high--energy
limit, while their scale dependence is not easily determined. However,
the leading nonlinear corrections to perturbative evolution, first
studied in Ref.\cite{Mueller:1985wy}, correspond to a suppression of
perturbative evolution due to gluon recombination. Therefore, to the
extent that saturation dynamics reproduces this suppression, we expect it
to be roughly compatible with the  effect we observe. 

Let us finally turn to the possible impact of the deviations we found
on LHC phenomenology. Indeed, if NLO DGLAP is affected by corrections
in the small $x$, small $Q^2$ part of the kinematic region which is
currently used for parton determination, one
must conclude that current parton sets, which use DGLAP evolution
throughout, will be affected by some bias due to lack of proper
inclusion of the necessary corrections to DGLAP evolution.
We will now  assess the size of this bias, and its impact on LHC observables.

In order to do this, we compute LHC observables using the PDFs
obtained with a variety of kinematic cuts: the PDFs obtained with more
restrictive cuts will be bias--free, while those obtained without cuts
might be biased. The difference between the observables computed in
these cases will thus give an indication of the possible amount of
bias.
On top of the cuts of the form Eq.~(\ref{eq:q2cutsat}) that we
discussed so far we will now also consider the impact on PDF
extraction of $Q^2$--independent cuts in $x$ of the form
\be
\label{mrstcut}
x_i \ge x_{\rm cut}.
\ee
These cuts have been considered
previously~\cite{Martin:2003sk,Huston:2005jm}; we will take the
same values of  $x_{\rm cut}$ as in these references. In comparison to the
cuts Eq.~(\ref{eq:q2cutsat})
one would expect them
to be equally effective in removing effects beyond NLO DGLAP, but to
also remove small $x$ data at larger $Q^2$, with consequent loss of accuracy.

\begin{table}
\centering
\begin{tabular}{|c|c|c|c|c|c|}\hline
$A_{cut}$ & $\sigma_{W^+}\mathcal B_{l^+ \nu_l}$ (nb) & 
 $\sigma_{W^-}\mathcal B_{l^- \nu_l}$ (nb) &
$\sigma_Z \mathcal B_{l^+ l^{-}}$ (nb) &
$\sigma_{gg->H}$ (pb)&
$\sigma_{t\bar t}$ (pb)\\
\hline
no cuts & 11.93 $\pm$ 0.30 & 8.43 $\pm$ 0.20 & 1.96 $\pm$ 0.04 & 36.6 $\pm$ 1.1 & 907 $\pm$ 23 \\
0.2 & 11.95 $\pm$ 0.33 & 8.46 $\pm$ 0.21 & 1.96 $\pm$ 0.04 & 36.6 $\pm$ 1.0 & 908 $\pm$ 22 \\
0.3 & 11.99 $\pm$ 0.41 & 8.50 $\pm$ 0.27 & 1.97 $\pm$ 0.06 & 36.5 $\pm$ 1.0 & 904 $\pm$ 28 \\
0.5 & 12.23 $\pm$ 0.37 & 8.62 $\pm$ 0.23 & 2.00 $\pm$ 0.05 & 36.8 $\pm$ 0.9 & 886 $\pm$ 29 \\
0.7 & 12.23 $\pm$ 0.36 & 8.68 $\pm$ 0.27 & 2.01 $\pm$ 0.05 & 37.0 $\pm$ 1.0 & 874 $\pm$ 35 \\
1.0 & 12.43 $\pm$ 0.46 & 8.71 $\pm$ 0.28 & 2.02 $\pm$ 0.05 & 37.4 $\pm$ 1.1 & 865 $\pm$ 34 \\
1.2 & 12.23 $\pm$ 0.46 & 8.66 $\pm$ 0.28 & 2.00 $\pm$ 0.06 & 37.6 $\pm$ 1.9 & 875 $\pm$ 35 \\
1.5 & 12.45 $\pm$ 0.44 & 8.72 $\pm$ 0.25 & 2.03 $\pm$ 0.05 & 37.7 $\pm$ 1.2 & 858 $\pm$ 36 \\ 
\hline
\end{tabular}
\caption{\small Results for LHC observables computed with PDFs
obtained from fits to reduced datasets based on
the `$A_{\cut}$' kinematic cuts Eq.~(\ref{eq:q2cutsat}).
\label{LHCobs}}
\end{table}

\begin{table}
\centering
\begin{tabular}{|l|c|c|c|c|c|}\hline
 & $\sigma_{W^+}\mathcal B_{l^+ \nu_l}$ (nb) & 
 $\sigma_{W^-}\mathcal B_{l^- \nu_l}$ (nb) &
$\sigma_Z \mathcal B_{l^+ l^{-}}$ (nb) &
$\sigma_{gg->H}$ (pb)&
$\sigma_{t\bar t}$ (pb)\\
\hline
no cuts & 11.93 $\pm$ 0.30 & 8.43 $\pm$ 0.20 & 1.96 $\pm$ 0.04 & 36.6 $\pm$
 1.1 & 907 $\pm$ 23 \\
$x>0.0002$  & 11.90 $\pm$ 0.32 & 8.39 $\pm$ 0.23 & 1.95 $\pm$ 0.04 &
36.5 $\pm$ 1.0 & 909 $\pm$ 24 \\
$x>0.001$ & 12.11 $\pm$ 0.36 & 8.55 $\pm$ 0.19 & 1.98 $\pm$ 0.04 & 37.0
$\pm$ 0.9 & 894 $\pm$ 25  \\
$x>0.0025$ & 12.75 $\pm$ 1.39 & 8.78 $\pm$ 0.62 & 2.05 $\pm$ 0.15 & 37.0 $\pm$ 1.1 & 886 $\pm$ 45 \\
$x>0.005$ & 12.46 $\pm$ 1.72 & 8.74 $\pm$ 0.93 & 2.03 $\pm$ 0.22 & 36.7 $\pm$ 0.9 &  877 $\pm$ 51 \\

\hline
\end{tabular}
\caption{\small Results for LHC observables computed with PDFs
obtained from fits to reduced datasets based on
the `$x_{\cut}$' kinematic cuts Eq.~(\ref{mrstcut}).
\label{LHCx}}
\end{table}

\begin{figure}
\epsfig{file=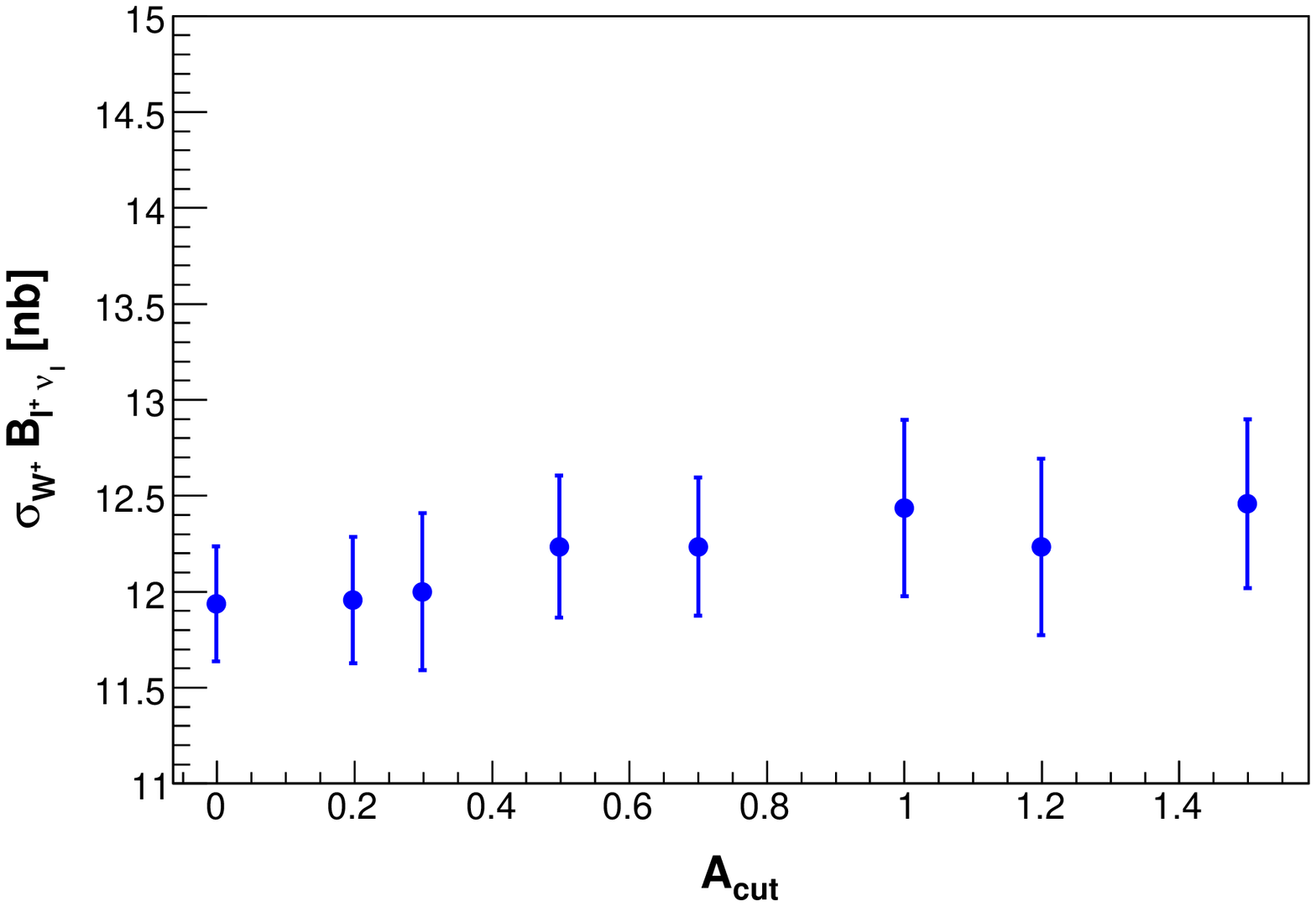,width=0.5\textwidth}
\epsfig{file=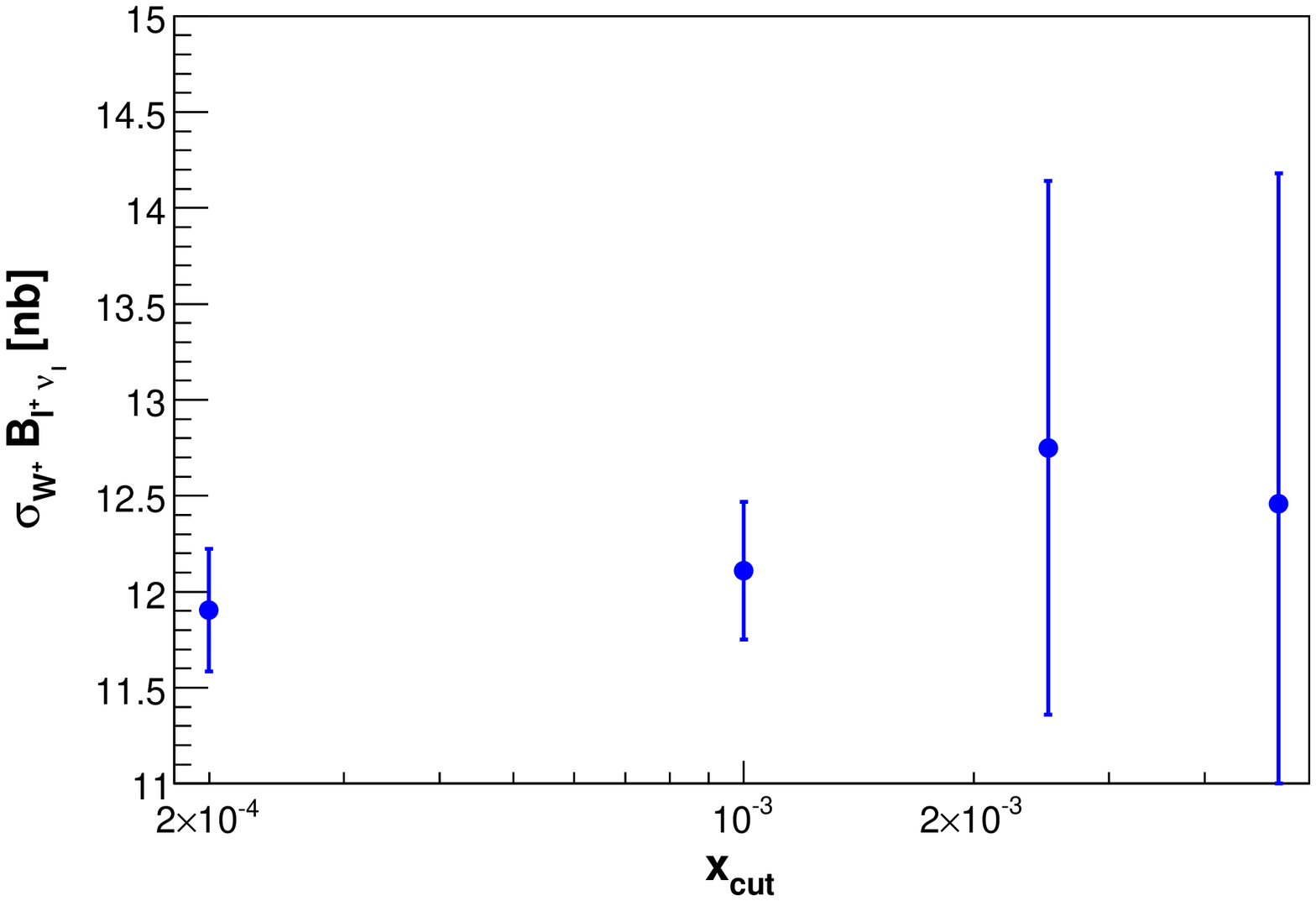,width=0.5\textwidth}
\caption{\small $W^+$ cross section for different cuts.
 Left: `$A_{\rm cut}$'-based fits. Right: `$x_{\rm cut}$'-based fits.}\label{Wfig}
\end{figure}

\begin{figure}
\epsfig{file=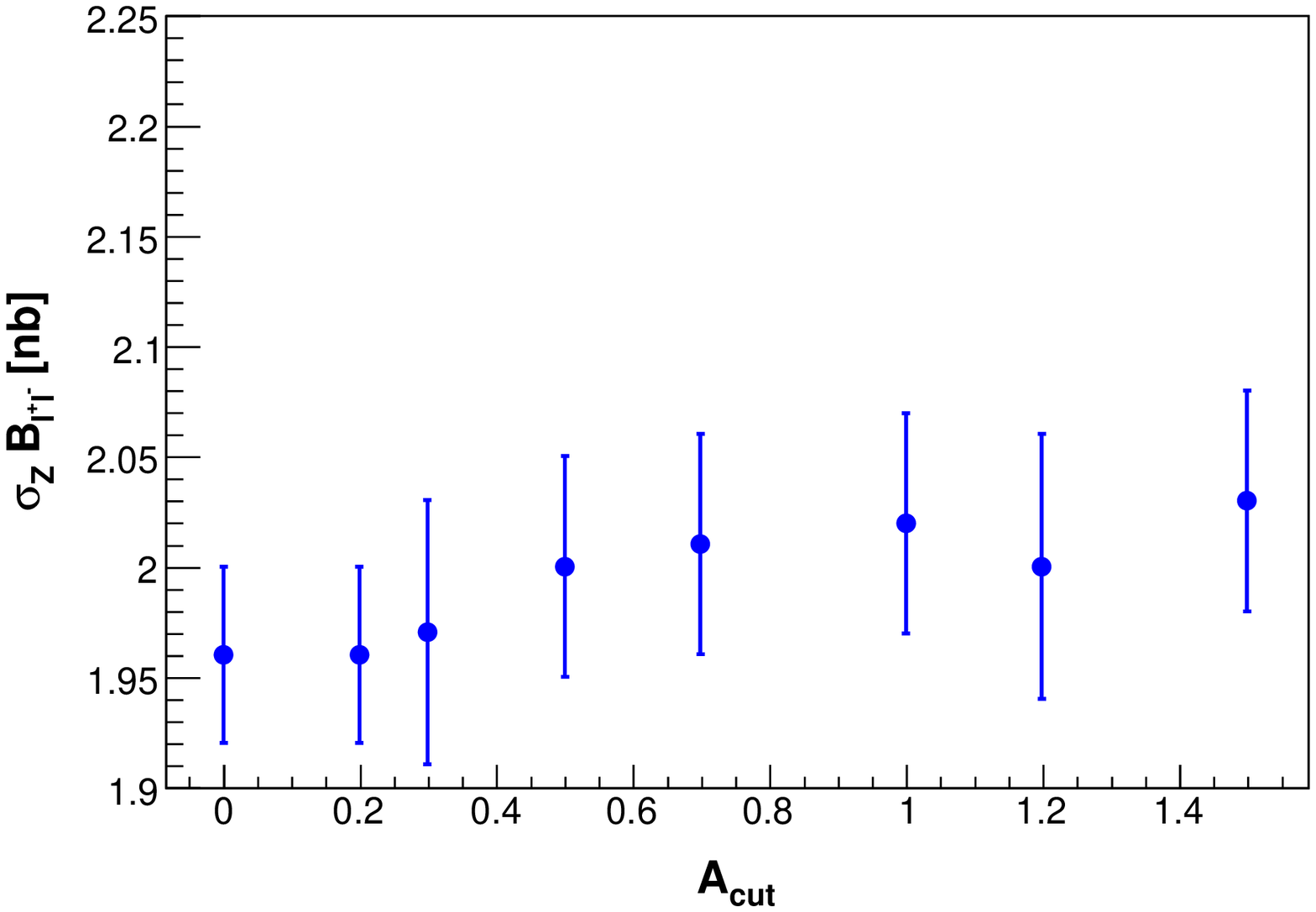,width=0.5\textwidth}
\epsfig{file=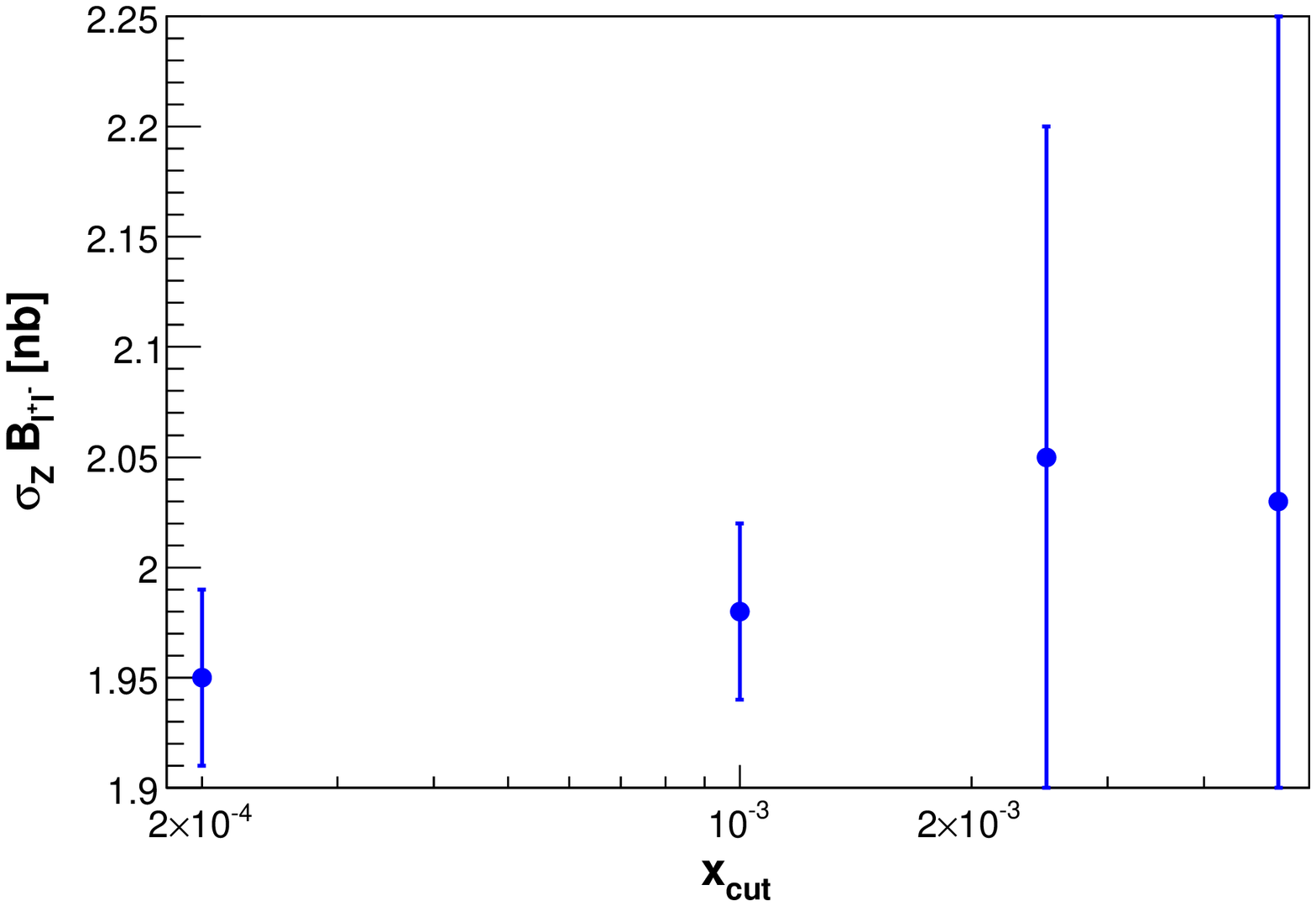,width=0.5\textwidth}
\caption{\small $Z$ cross section for different cuts.
Left: `$A_{\rm cut}$'-based fits. Right: `$x_{\rm cut}$'-based fits.}
\label{Zfig}
\end{figure}

\begin{figure}
\epsfig{file=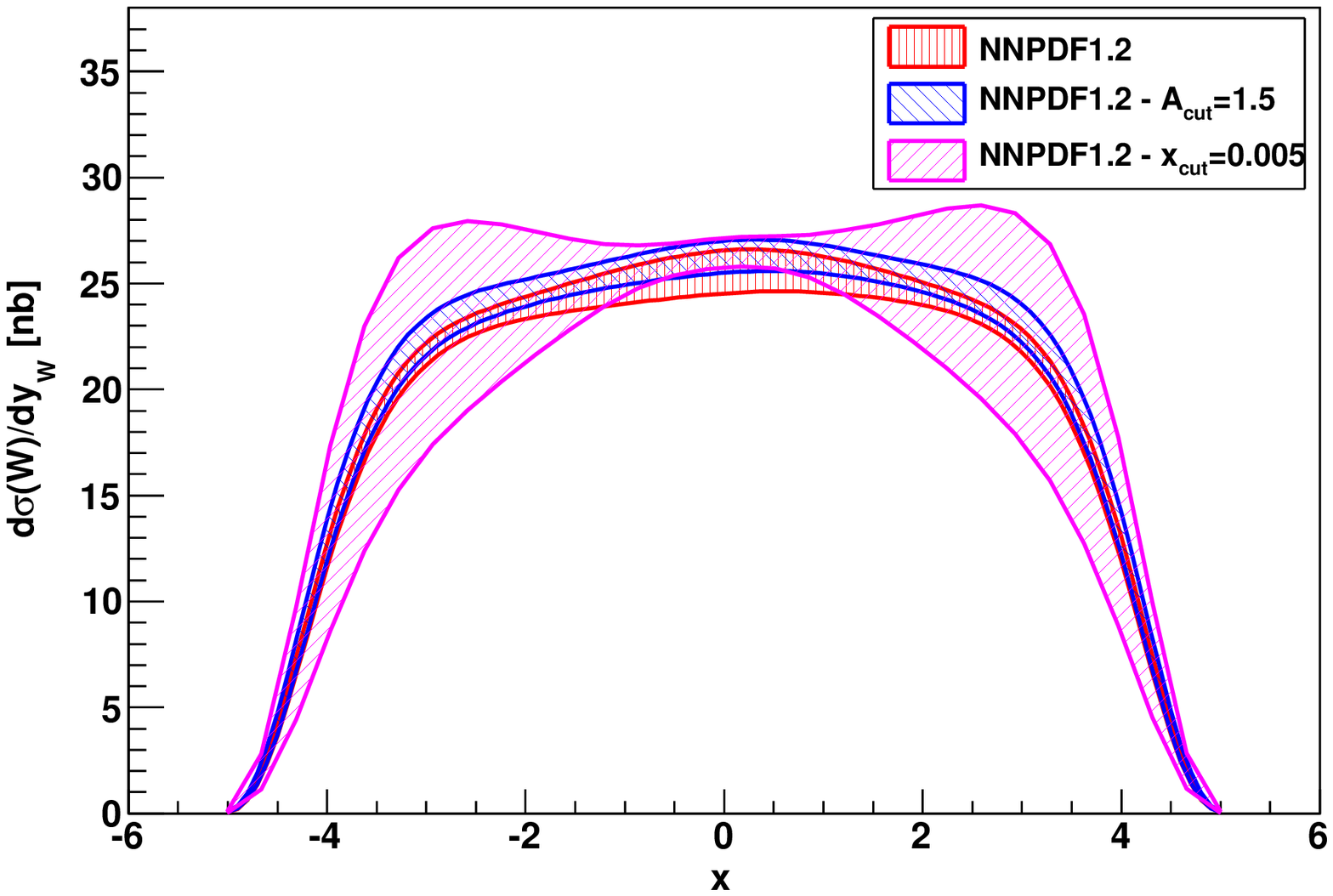,width=0.5\textwidth}
\epsfig{file=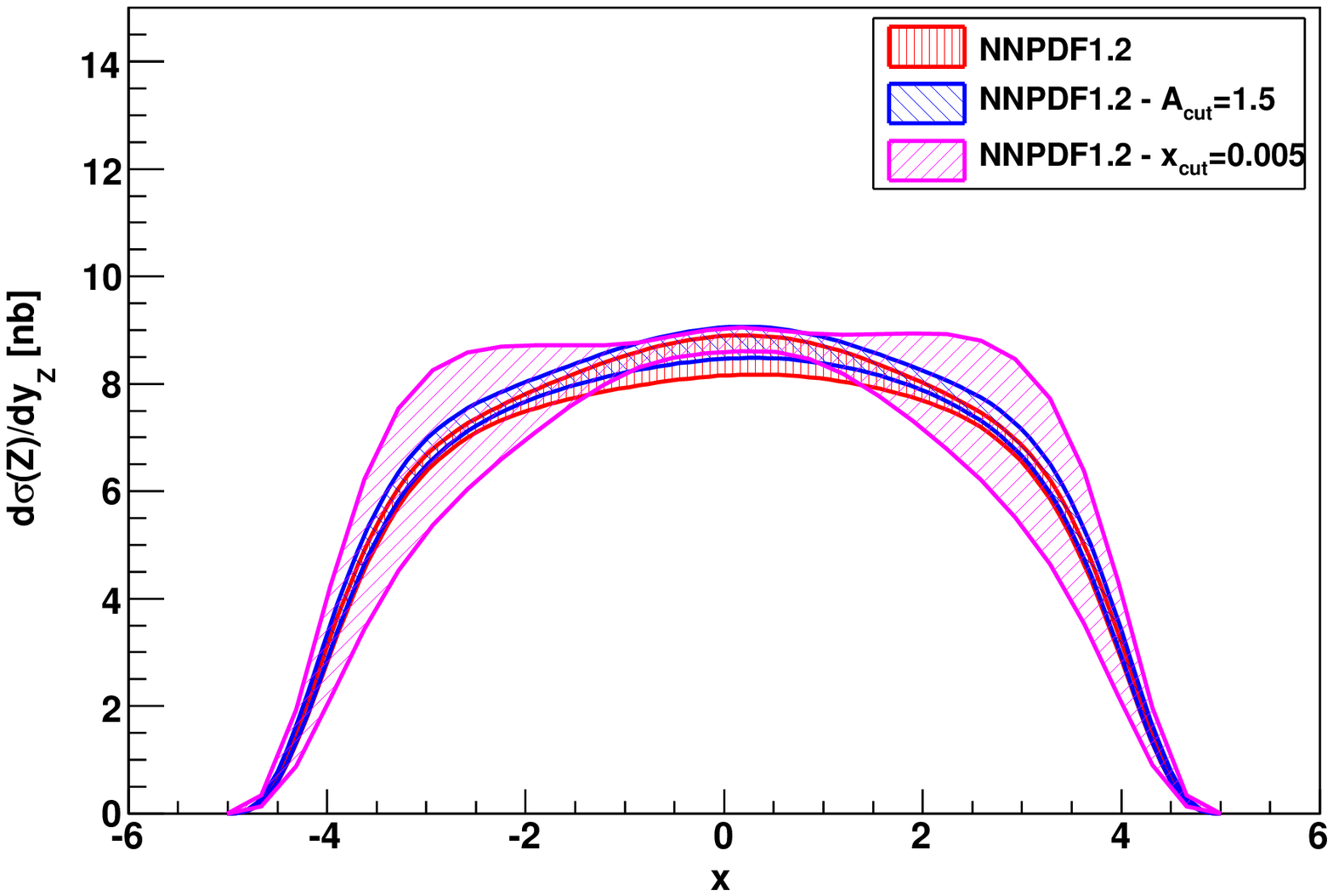,width=0.5\textwidth}
\caption{\small The $W$ (left) and $Z$ (right) rapidity distributions
  at the LHC computed with different cuts.}\label{WZrapfig}
\end{figure}

Using the PDFs obtained in all these ways, 
we have computed for LHC kinematics
(at 14~TeV) the $W^{\pm}$, $Z^0$,
$t\bar{t}$ and $gg\to H$ total NLO cross sections (using the MCFM
code~\cite{mcfm}) and rapidity distributions (using a
code~\cite{vicini} benchmarked against MCFM).
Results for total cross sections
are summarized in Tables~\ref{LHCobs} and~\ref{LHCx} 
and in a graphical way in Figs.~\ref{Wfig} and~\ref{Zfig}, while the
rapidity distributions are shown in Fig.~\ref{WZrapfig}.
Clearly, central values for all observables with the exception of
Higgs production change significantly as the cut is raised. 
However, PDF uncertainties
increase accordingly, in such a  
way that the observables from the fits from reduced datasets
 remain always
compatible with the observables from the reference fit. The greater
stability of the Higgs production cross section
can be understood as a consequence of 
the fact that it depends essentially on the large-$x$
gluon, which is essentially unaffected by the cuts, while
the other observables depend also on the small $x$ behaviour of PDFs.
Interestingly, as the cut is raised the shift in 
central values of 
observables (such as $W$ and $Z$ production) is
similar to that induced by the inclusion of 
charm mass effects~\cite{Tung:2006tb,Nadolsky:2008zw}, consistent with
our observation that these effects could explain part of the
discrepancy with NLO DGLAP that we find.
% ----------------------------------

If we compare 
`$A_{\cut}$'  fits (based on cuts Eq.~(\ref{eq:q2cutsat})) 
and `$x_{\cut}$' fits (based on cuts Eq.~(\ref{mrstcut})), we see that
the loss of accuracy in the latter
case is considerably larger. Therefore, an optimal choice of
`conservative'~\cite{Martin:2003sk} partons, which minimizes the impact
of possible non-standard effects while maximizing the use of available
information, could be based on the choice of a cut of the form
Eq.~(\ref{eq:q2cutsat}). Also, due to this large
increase in the PDF uncertainty, we do not have evidence of inconsistencies
between PDFs obtained from
`$x_{\cut}$' fits and the reference, in partial disagreement with the
conclusion of Ref.~\cite{Martin:2003sk}, though perhaps this
conclusion might also change if one used the more flexible parton
parametrization and dynamical tolerance methods recently introduced by
the MSTW collaboration~\cite{Martin:2009iq}.
% ---------------------------------

Our general conclusion is that the possible distortion of NLO PDF sets
induced by deviations from NLO DGLAP in the small $x$ and $Q^2$ region
of the  HERA data has an effect on the NLO determination of inclusive
LHC observables which is at the percent level. However, the full impact on LHC
observables of  the dynamics which might cause the deviations (such as
small $x$ resummation) could only be assessed by a dedicated study
which goes beyond the scope of this work.

In summary, we have  implemented a strategy to single out deviations from
fixed order DGLAP evolution in a QCD analysis of
inclusive experimental data. We have applied this strategy to a
PDF analysis based on the NNPDF framework, and studied the
effects of various kinematical cuts in HERA
data. The use of several statistical indicators shows clear evidence
for deviations between the scale dependence observed in the data and
that predicted by NLO DGLAP evolution in the small $x$ and $Q^2$
region, and various checks indicate that the effect is unlikely
to be due to the fitting methodology because it is a local feature of
the evolution observed in the region under consideration.
Whereas part of these deviations could be removed by a more
refined treatment of the charm threshold, the effect is likely to be
due to either the resummation of small $x$  perturbative corrections, or
power--suppressed corrections to perturbative evolution. It cannot be
explained by invoking NNLO terms, which go in the wrong direction and
would  make the discrepancy worse.
We then have shown that even if these effects induce a distortion of
available PDF set, the impact of this distortion on the NLO
computation of standard candle
LHC observables
is  small in comparison to 
current PDF uncertainties. It might however become more significant in
future precision studies, especially if deep-inelastic scattering data
at
higher energy were available, such as those which might be obtained at
a future electron-hadron collider based on the LHC
(LHeC)~\cite{Dainton:2006wd,Rojo:2009ut}. More interestingly, if the
cause of these effects were established with certainty, it might require
a nontrivial reassessment of the determination of LHC observables: for
example, by a systematic inclusion  of small $x$ resummation effects.
\medskip

{\bf \large Acknowledgements}\\
We thank all the members of the NNPDF collaboration, which has
developed the PDF fitting methodology and code on which this study is
based.  
We acknowledge discussions with  J.~L.~Albacete,
N.~Armesto and G.~Milhano. 
This work was partly supported 
by the European network HEPTOOLS under contract
MRTN-CT-2006-035505.

\end{document}